\documentclass[a4paper,fleqn,usenatbib]{mnras}

\usepackage{newtxtext,newtxmath}
\usepackage{mathptmx}
\usepackage{longtable,lscape}

\usepackage[T1]{fontenc}
\usepackage{ae,aecompl}
\usepackage{supertabular}
\usepackage{longtable}

\usepackage{graphicx}	% Including figure files
\usepackage{amsmath}	% Advanced maths commands
\usepackage{amssymb}	% Extra maths symbols
\usepackage[none]{hyphenat}

\title[Halo wide binaries from SDSS]{A Distant Sample of Halo Wide Binaries from SDSS\thanks{Based in part on observations collected through CNTAC proposals CN-2015A-0611, CN-2015B-0605, ESO proposals 095.D-0563(A) 095.D-0563(B) and NOAO proposal CN-2015A-0080}}

\author[J. Coronado et al.]{Johanna Coronado,$^{1,2,3}$\thanks{E-mail: coronado@mpia.de} Mar\'ia Paz Sep\'ulveda,$^{1,2}$ Andrew Gould$^{3,4,5}$ and\newauthor Julio Chanam\'e$^{1,2}$ 
\\
$^{1}$Instituto de Astrof\'isica, Pontificia Universidad Cat\'olica de Chile, Av. Vicu\~na Mackenna 4860, 7820436 Macul, Santiago, Chile\\
$^{2}$Millennium Institute of Astrophysics, Santiago, Chile\\
$^{3}$Max-Planck-Insitut f\"ur Astronomie, K\"oningstuhl 17, D-69117 Heidelberg, Germany\\
$^{4}$Korea Astronomy and Space Science Institute, Daejon 34055, Korea\\
$^{5}$Department of Astronomy, Ohio State University, 140 W. 18th Ave., Columbus, OH43210, USA\\
}

\date{Accepted XXX. Received YYY; in original form ZZZ}

\pubyear{2018}

\begin{document}
\label{firstpage}
\pagerange{\pageref{firstpage}--\pageref{lastpage}}
\maketitle

\begin{abstract}
Samples of reliably identified halo wide binaries are scarce.  If reasonably 
free from selection effects and with a small degree of contamination by 
chance alignments, these wide binaries become a powerful dynamical tool, 
having provided one of the very few experiments capable of constraining the nature of dark matter in the Milky Way halo.  Currently, however, the best available sample of halo wide binaries is confined to the solar neighborhood, and is plagued by small number statistics at the widest separations.  We present the results of a program aimed to probe the wide binary population of the Galactic halo at significantly larger distances, and which informs future searches that could improve the statistics by orders of magnitude. Halo stars were taken from the Sloan Digital Sky Survey after analysing the Galactic orbits of stars in the reduced proper motion diagram.  We then select candidate binaries by searching for pairs with small differences in proper motion and small projected separation on the sky.  Using medium-resolution spectroscopy, a subsample of candidates is validated via radial velocities, finding that about 68\% of candidate pairs up to 20$^{\prime\prime}$ separation can be considered genuine halo wide binaries, to the limits of the available data.
Precise parallaxes from \textit{Gaia} confirm that most of our selected pairs have their components at the same distances, independently confirming the robustness of our selection method.
These results should prove valuable to guide the optimal assembly of larger catalogs of halo wide binaries from upcoming large databases, e.g., \textit{Gaia} and LSST.

\end{abstract}

\begin{keywords}
catalogues -- astrometry -- binaries: general -- stars: kinematics and dynamics
\end{keywords}

%%%%%%%%%%%%%%%%%%%%%%%%%%%%%%%%%%%%%%%%%%%%%%%%%%

%%%%%%%%%%%%%%%%% BODY OF PAPER %%%%%%%%%%%%%%%%%%

\section{Introduction}

Observations have established that multiplicity is the principal product of star formation \citep{Reipurth} and multiple systems of two or more bodies seem to be the norm at different stellar evolutionary stages and throughout stellar types. In consequence, stars in binary systems provide valuable information and influence almost all the branches in astronomy. However, most of this knowledge has come from the study of binary stars in close orbits, while the population of binaries with semi-major axes $a\gtrsim$ 100 AU (often referred to as wide binaries) remains poorly explored, comparatively speaking \citep{ch07}. 

Wide binaries have low binding energies and large collisional cross sections, which makes them easily disrupted by any inhomogeneity of the underlying gravitational potential (e.g., passing stars, molecular clouds, spiral arms, etc.) large scale tides, or even Massive Compact Halo Objects (MACHOs). If these encounters are frequent during the lifetime of the wide binaries in our Galaxy, one would expect the distribution of binaries to be depleted for the largest separations.
Therefore, wide binaries can be used as probes to establish the properties of such perturbers. For example, they have been used to place constraints on the amount of halo dark matter composed of compact objects above some mass threshold \citep{bahcall, yoo04,quinn1,allen,penarrubia}.

Additionally, the distribution of their semi-major axis might contain information about the dynamical history of the Galaxy. Wide binaries are also important for understanding the birth environment of stars. For example, their formation has been studied through unfolding of triple systems \citep{Reipurth2} and during star cluster dissolution \citep{Kouwenhoven}. Also, \citet{moeckel} studied their formation in expanding clusters. \citet{jiang} studied the orbital evolution of wide binaries in the solar neighborhood including the effects of passing stars and the Galactic tidal field. Finally, wide binaries are also useful to put constraints for gyrochronology \citep[see][]{chaname,godoy}

Starting from the revised New Luyten Two Tenths catalog \citep[rNLTT;][]{sal03}, \citet{cg04}, hereafter CG04, assembled the first sample of halo wide binaries large enough for reliable population analyses. They cataloged 116 halo wide binaries with angular separations up to $\Delta\theta$ = 900'' (projected separations of $\sim$1~pc). Using this sample, \citet{yoo04} set constraints on the mass and density of compact Galactic halo dark matter, excluding MACHOs with masses M $>$ 43M$_{\odot}$, at the local standard halo density. On the other hand, MACHOs with smaller masses ($M\lesssim 30M_\odot$) had already been excluded by micro-lensing surveys \citep{udalski,alcock1,alcock2,Afonso2003}. 
Consequently, the combination of both experiments left only a small window for a Galactic halo composed entirely of MACHOs. 

Nevertheless, the \citet{yoo04} constraints rely heavily on the genuineness of the four widest binaries of the CG04 sample. Using radial velocity (RV) measurements, \citet{quinn1} showed that one of these four pairs is most likely a spurious one. Omitting this pair, the constraints on MACHO masses are substantially relaxed, increasing the unconstrained window to $\approx$30-500~M$_{\odot}$. This sensitivity clearly means that larger samples of wide binaries with as little contamination as possible are needed to put stronger constraints on MACHOs.
Recently, \citep{brandt} put new constraints on MACHOs of M $\gtrsim$ 5M$_{\odot}$ as the main component of dark matter from the ultra-faint dwarf galaxy Eridanus II using the survival of one of its star clusters. However, this result is based on a single object, and, moreover, the subject is of such importance that it is desirable to contrast its validity against experiments from independent directions. Therefore, the test provided by wide halo binaries remains of high importance.
Significant effort has been put during the past decade to search for halo wide binaries.
For instance, \citet{quinn&smith} searched for halo wide binaries in the stripe 82 of the Sloan Digital Sky Survey (SDSS) and found 16 pairs of candidates with projected separations between 0.01 and 0.25 pc. Although this sample on its own it is not enough to improve constraints on dark matter, it is a useful addition to the CG04 sample of wide binaries.
More recently, \citet{allen14} compiled a list of 210 halo systems by joining existing catalogs.
However, their source catalogs were assembled independently, with different selection biases and degrees of completeness, most importantly as a function of semi-major axis. Therefore, since what matters for dynamical purposes is precisely the semi-major axis distribution of the population of binaries, the act of combining different catalogs does not produce a clean sample to draw robust conclusions regarding dark matter constraints.

Having precise astrometric data of a large number of stars is of paramount importance to assemble a significant sample of wide binaries belonging to our Galactic halo.
This will be possible with the upcoming results from \textit{Gaia's} second data release (DR2), although some effort has already been done towards that goal using the Tycho-Gaia astrometric solution (TGAS) catalogue from \textit{Gaia's} first data release  \citep[DR1;][]{Oh,andrews1,andrews2}. 
However, this release is mostly confined to the solar neighborhood, containing almost no halo population. 
The big challenge for these very large astrometric catalogs is then to be able to distinguish the truly bound systems from among the large number of false pairs that will inevitably be present in such a large sample of stars.
With \textit{Gaia's} DR2 we will have access to better astrometric information and therefore we will be able to better discriminate genuine wide binaries from spurious associations. Nevertheless, since contamination grows rapidly with separation \citep[see][]{andrews1,andrews2} the very widest samples will need radial velocity confirmation, albeit not necessarily selection. Hence, to assemble a catalog of highly reliable (i.e., genuine) halo wide binaries, which as demonstrated above is fundamental for deriving solid dark matter limits, the radial component of the space velocities is ultimately needed for the candidates that satisfy the proper motion, parallax, and/or photometry criteria.

In this paper, we present the results of a small-scale program for selecting reliable, genuine halo wide binaries using the SDSS, accounting properly for contamination from unassociated pairs. The outline of this work is as follows: In section 2 we present the selection of the candidate halo wide binaries, in section 3 the observations and data reduction, in section 4 the analysis of candidates and in section 5 the conclusions.

\section{Selecting the Sample}
\label{sec:rpm}

We selected our sample using proper motions and photometry from the SDSS, which is an imaging and spectroscopic survey that covers a large area of the sky in five optical bands \textit{u,g,r,i,z}, to a depth of g$\sim$ 23. The survey uses a dedicated 2.5 m telescope equipped with a large-format mosaic CCD camera to image the sky, and two digital spectrographs to obtain the spectra of about 1 million galaxies and 100,000 quasars selected from the imaging data \citep{york}. 
SDSS astrometry is calibrated using observations by an array of astrometric CCDs in the imaging camera \citep{york}. Although it is a single-epoch survey, it can be combined with older photometric surveys in order to produce two epochs, necessary to compute proper motions of its stars. 
This was accomplished by \citet{gould2004} and later by \citet{munn04} who derived proper motions by combining SDSS astrometry with USNO-B, with statistical errors of roughly 3-3.5 mas/yr. 
We use Data Release (DR) 9, which has the advantage of correcting errors in proper motions, previously present for DR7 stars
at low Galactic latitudes. Proper motions in SDSS are available through the \textit{proper motions} table in CasJobs\footnote{\url{http://skyserver.sdss.org/casjobs/}}, which provides a platform to query and access a large data collection in SDSS. 
\subsection{Selection of halo stars from SDSS}
\label{sec:selection_of_pairs}

Using the aforementioned query, we choose stars from DR9 with SDSS $r$-band magnitudes $14<$\textit{r}$<20$ and proper motions $|\vec\mu|>$ 30 mas/yr.
This latter imposed criterion over the proper motions significantly decreases the density of stars, especially more distant ones, which in turn considerably decreases the probability of chance alignments (i.e. stars move together in the sky, but that are not physically related).

For the query, psf magnitudes were chosen, with their corresponding extinction coefficients. In order to have reliable photometry we follow \citet{ses08} to select real sources. From this query \textit{2,767,380} stars were obtained, with errors in photometry less than 0.05 mag.

The selection for this catalog so far is comprised of halo and disc stars, thus the next step is to separate both populations in order to select just halo stars. With only proper motion and photometry information, the Reduced Proper Motion (RPM) diagram can be use to distinguish between halo and disc stars.
\citet{sal03} first accomplished the task of constructing a clean RPM sample by improving the astrometry and photometry of the high proper motion ($|\vec\mu|\geq$ 180 mas/yr) stars from the New Luyten Two Tenths (NLTT) catalog. 

The RPM diagram is a powerful tool to separate halo from disc stars because although these stars have similar absolute magnitudes, they have very different kinematics, and as a consequence, the faster-moving halo stars appear offset from the disc stars in an RPM diagram \citep{smith09}. They are also fainter (because in general more distant) and bluer, which also helps in the separation. The V-band RPM is defined as Eq.~\ref{eq:v_rpm}:
\begin{equation}
\label{eq:v_rpm}
\mbox{V}_{\mbox{\small{RPM}}} = \mbox{V} + 5\mbox{log} |\vec\mu|,
\end{equation}
and then plotted against a color, where $\mu$ corresponds to the proper motion. In this case, considering the SDSS g band, we define the RPM as:
\begin{equation}
 \mbox{g}_{\mbox{\small{RPM}}} = \mbox{g} + 5\mbox{log} |\vec\mu|  -1.47 |\mbox{sin(b)}|,
 \label{eq:g_rpm}
\end{equation}
where $|$sin(b)$|$ is a correction factor that corrects for the vertical offset, introduced by different Galactic latitudes \citep{sal03}.

\begin{figure}
\begin{tabular}{c}
\includegraphics[width=\columnwidth]{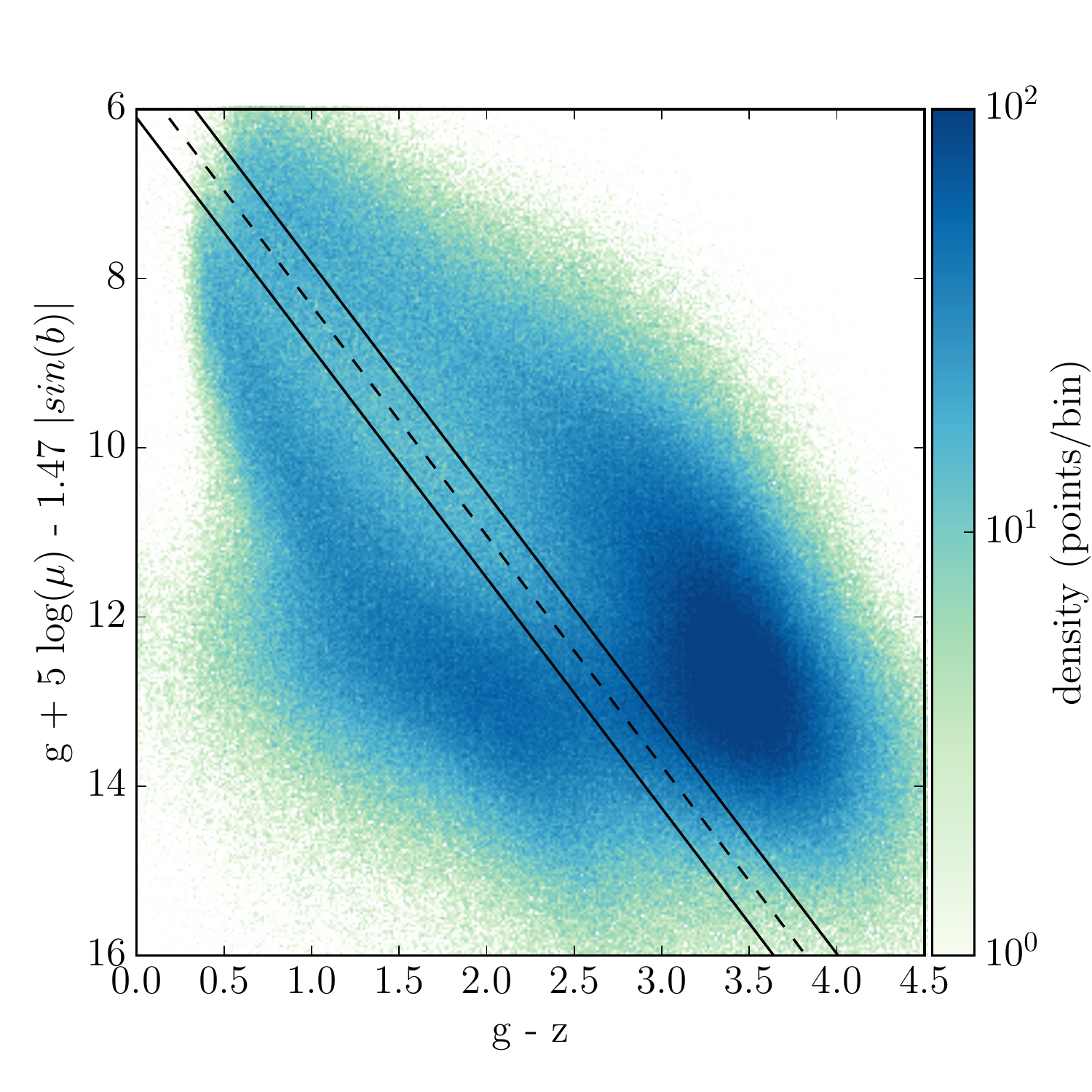}\\
\includegraphics[width=\columnwidth]{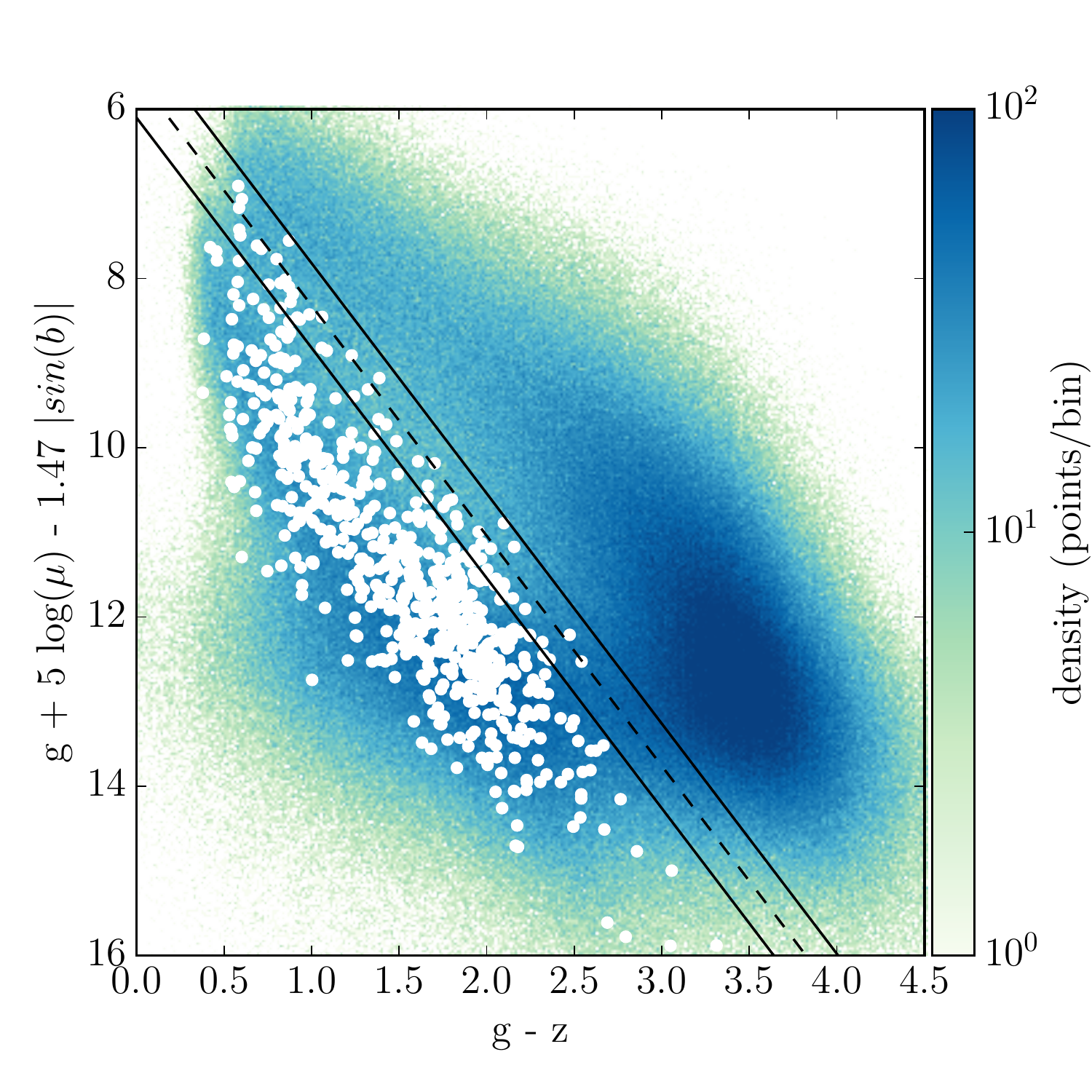}
\end{tabular}
\caption{Reduced proper motion (RPM) diagram of the stars in our sample with proper motions of $|\vec\mu| > 30$ mas/yr.
This diagram allows us to discriminate between disc and halo populations that are located to the right and left side of the confidence lines, respectively. The white circles show our common proper motion pairs that satisfy our criteria $\mu$/$\Delta\mu > 10$ and 5''$\leq \Delta\theta \leq$ 200''. The upper solid line is a relaxed limit separating both populations, the dashed line is a more confident limit and the third solid line, the one closest to the halo, is a confident separator between halo and disc.}
    \label{fig:rpm_all_before}
\end{figure}

In Fig.~\ref{fig:rpm_all_before} we observe by simple inspection a separation between a more dense and populated disc extending from 3.0$< g-z<$ 4.5 and the halo extending from $\sim 0.5<g-z<2.5$. As a first approach, the position and slope of the separating lines in this plot can be arbitrarily placed based on a visual analysis of the location of the halo and disc tracks in the diagram.

However, as it is crucial that our sample contains only halo stars, we want to determine a more robust limit.
With this aim, we performed an analysis of the RPM diagram using radial velocity measurements obtained from SEGUE\footnote{Sloan Extension for Galactic Understanding and Exploration}. This analysis was done by selecting only halo stars ($\sim$ 400,000 stars) from the RPM diagram, and we found that only 30,000 of these had radial velocity measurements. Also, from this query, we requested the full parameter space information available for these stars to calculate the orbital parameters such as eccentricity, maximum height with respect to the disc plane ($z_{max}$) and maximum and minimum distance to the Galactic Centre. 

\begin{figure}
\centering
	\includegraphics[width=\columnwidth]{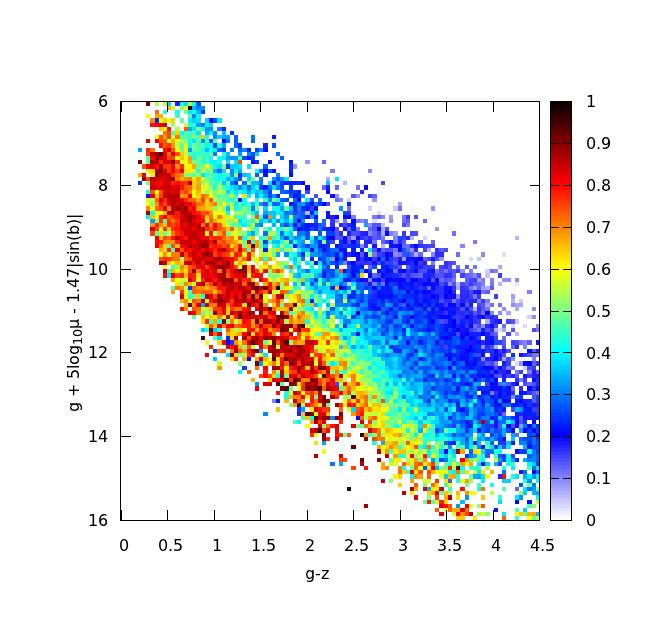}
    \caption{RPM color coded by eccentricity. We can see a clear separation between stars in eccentric orbits, belonging to the halo, and stars with eccentricities closer to circular orbits, typical of disc stars.}
    \label{fig:RPM_ecc}
\end{figure}

Halo stars move on random, more eccentric orbits, therefore an analysis on the obtained orbits allowed to separate more clearly between halo and disc stars, as illustrated in Fig.~\ref{fig:RPM_ecc}. Here, we color code the RPM by eccentricity and we clearly see a separation between halo and disc stars. This is the final criterion we apply to separate disc from halo in the RPM diagram and to place the dashed and solid lines in Fig.~\ref{fig:rpm_all_before}. 
In this figure the upper solid line is a relaxed limit separating both populations, the dashed line is a more confident limit and the third solid line, the one closest to the halo, is a confident separator between halo and disc. 

\subsection{Selection of common proper motion pairs}
\label{sec:sel_pairs}
Once the catalog of halo and disc stars is assembled, we select halo stars applying the criterion described in \S \ref{sec:rpm}. The next selection criterion is to consider pairs with angular separations between
5$^{\prime\prime}$ $\leq \Delta\theta \leq$ 200$^{\prime\prime}$. A lower limit of 5$^{\prime\prime}$ is considered because of the limits in photometry, at smaller angular separations it is harder to distinguish between stars and blending can appear. Also, this avoids potential confusion that may be inherent to our source proper motions, which were obtained by comparing the SDSS and USNO-B epochs \citep{gould2004,munn04}. The upper limit is considered because at larger angular separations, more contamination of unassociated stars is present. 
Since our sample is composed by stars from the Galactic halo, we expect our most distant pairs to be a few kpc from the Sun. 
An angular separation of 200$^{\prime\prime}$ corresponds to a projected separation of $\sim$1 pc at 1 kpc, and thus this upper limit still allows for very wide pairs, in principle, to be selected.

A criterion in proper motion is also applied, selecting small differences in proper motion $\Delta \vec\mu$ for the pairs. We choose candidates with $\mu$/$\Delta\mu > 10$. 
Larger values of $\mu$/$\Delta\mu$ could imply several possibilities; that the difference in proper motion of the halo wide binary candidate is very small, that these move with very high proper motions, or a combination of both. 
This sets strong constraints on the selection of the candidates given that if both stars move with high proper motions the probability that these stars are not related is very low, and in the case that both have very similar proper motions (hence, smaller differences in $\mu$), then the probability that both are related is very high. Therefore, for all possibilities, the larger the quantity, the more likely the real association between the two stars. Indeed, the high rate of success after radial velocity vetting (Section 4) demonstrates the reliability of this quantity for the selection of good candidates. The specific threshold was chosen because larger values of  $\mu$/$\Delta\mu$ leave too few candidates and smaller values increase the possibility of contamination.

\begin{figure}
    \centering
        \includegraphics[width=\columnwidth]{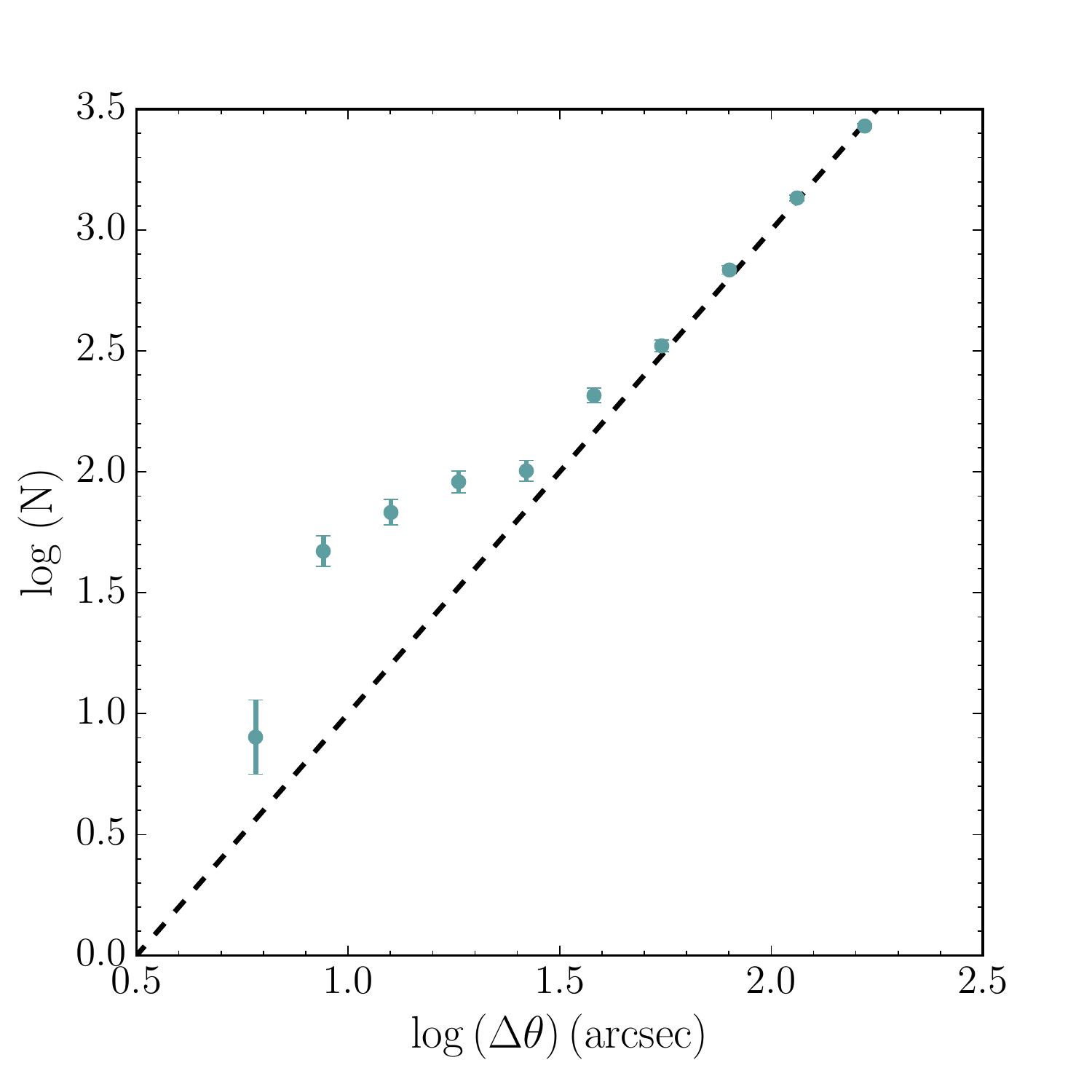}
        \caption{Distribution of angular separations for our initial sample of halo wide binary candidates ($\sim$ 5000 pairs). This distribution shows the number of pairs in equal logarithmic bins of angular separation and it is obtained by just counting pairs. The error bars show Poisson uncertainties. We note an excess of pairs at close angular separations, indicating the presence of true bound systems. The dashed line has slope 2, since random association of stars increase $\propto\theta^2$}
\label{fig:ang_sep1}
\end{figure}

Applying the criteria in proper motion and angular separation, selecting pairs that satisfied this condition the catalog is now composed of $\sim$ 5000 pairs.
In Fig.~\ref{fig:ang_sep1} we present the distribution of these pairs in equal logarithmic bins of angular separation.  
We clearly see an excess with respect to the random distribution (dashed line) of pairs at close angular separations $5^{\prime\prime}<\theta<32^{\prime\prime}$, which already suggests the presence of genuine binaries. 
At larger separations we see that the number of pairs steeply increases, indicating the presence of contamination by chance alignments, as was already noted by \citet{ses08} and \citet{quinn&smith} who also analysed the angular distribution of pairs in SDSS. 

\subsection{Assessing the \textit{binarity} of our candidates}
\label{sec:binarity}

If we want to determine the \textit{binarity} of our candidates, i.e. the confidence to which we can say a system is a genuine binary, we need to 
establish a method to disentangle the true bounded systems from line-of-sight chance alignments. To this end, we first explore in detail the distribution in angular separation of our entire initial catalog $\sim$ 5000 pairs as showed in Fig.~\ref{fig:ang_sep1} without including radial velocity information. 

\begin{figure}
\begin{tabular}{c}
\includegraphics[width=\columnwidth]{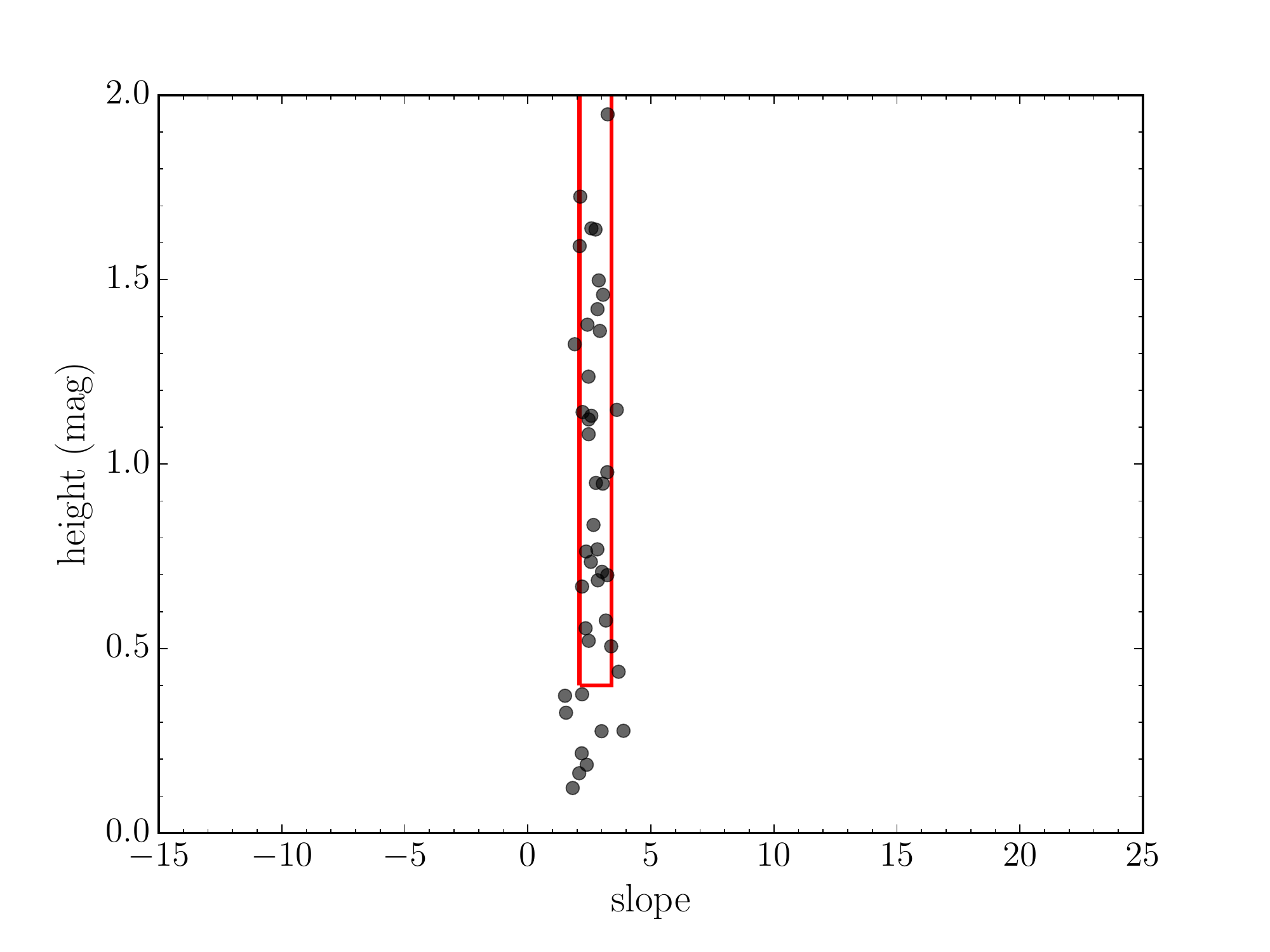}\\
\includegraphics[width=\columnwidth]{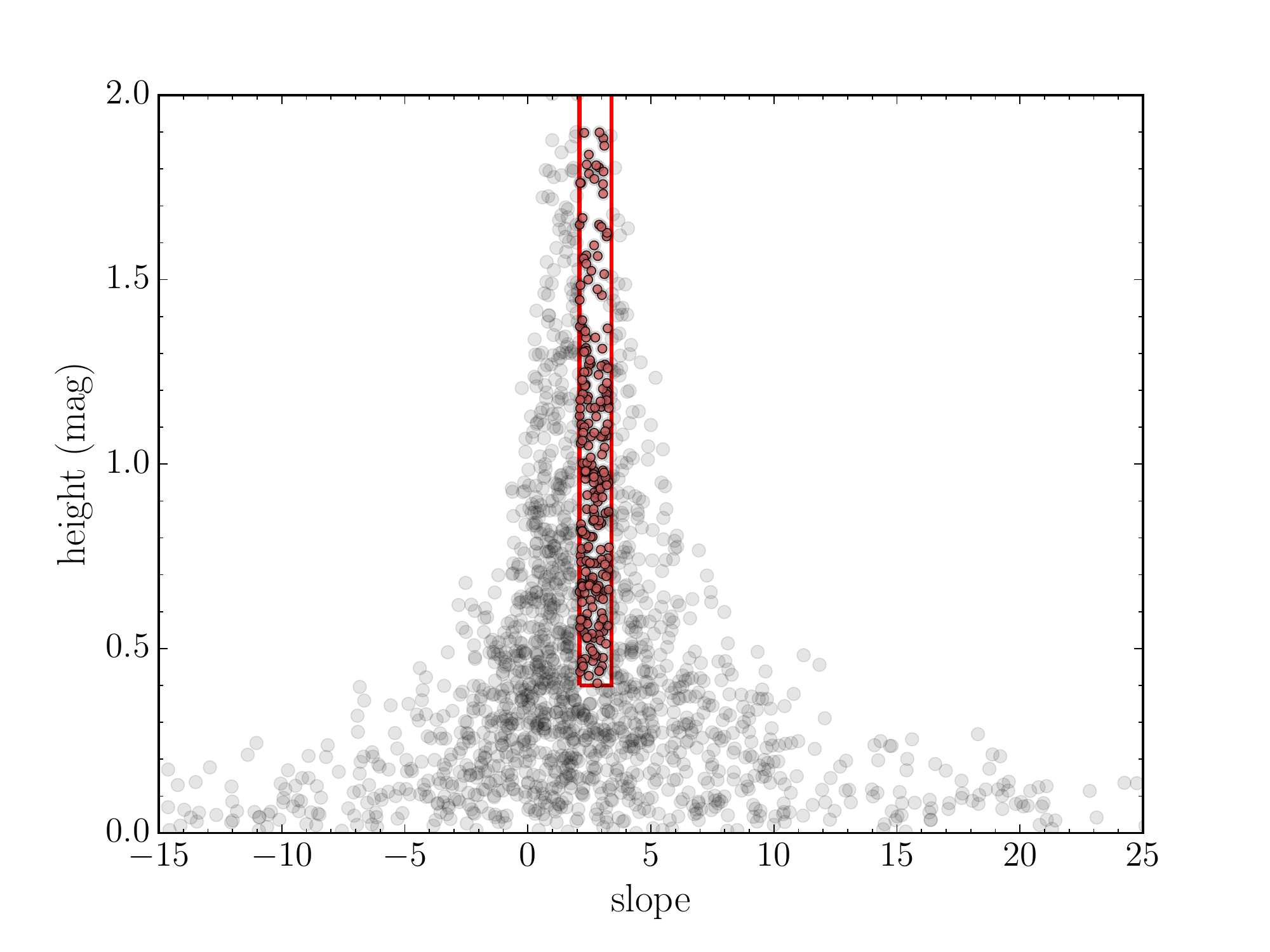}
\end{tabular}
\caption{
Height and slope of the \textit{isotach} connecting candidate pairs in the RPM diagram. In the upper panel we show pairs from the third bin in Fig.~\ref{fig:ang_sep1}, while the lower panel shows the pairs that fall in the last bin of the same figure. Both samples represent two different regimes -- where we expect to find mostly genuine binaries and the latter where we are dominated by contamination. The red rectangle shows the region of this plot where we should have well behaved pairs connected by a parallel line with respect to the halo track in the RPM.}
\label{fig:slope1}
\end{figure}

As we can see from this figure, the last two bins lie on top of the dotted line that shows how the contamination fraction increases with angular separation. 
The last two bins of Fig.~\ref{fig:ang_sep1} follow closely the behaviour of random pairs, making the identification of true pairs impossible. 
Therefore, we can use the last bin to account for contamination from chance alignments in our catalog. 
In order to do this we plot in the lower panel of Fig.~\ref{fig:slope1} the slope and height from the \textit{isotachs} connecting pairs in the RPM for the last bin in Fig.~\ref{fig:ang_sep1}. We define the height as the difference in the $y$-axis of each pair in the RPM diagram. 

\begin{figure}
    \centering
        \includegraphics[width=\columnwidth]{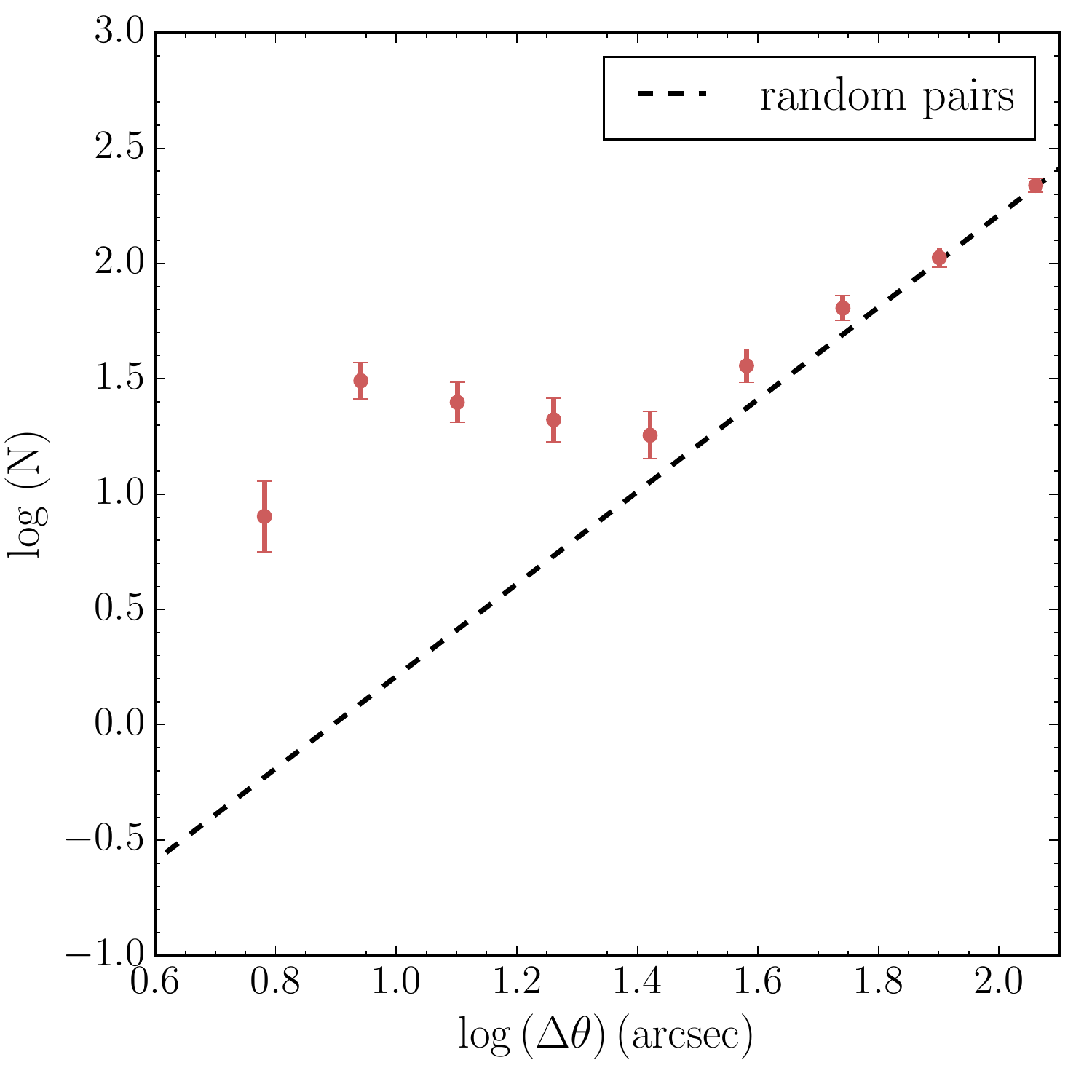}
        \caption{Distribution of angular separations as seen in Fig.~\ref{fig:ang_sep1} but now counting the pairs that fall inside the region defined in Fig.~\ref{fig:slope1} for each bin. The dashed line has the same slope as Fig.~\ref{fig:slope1}, but with a different zero point. This was was obtained by using the fraction of pairs from the last bin that fall into the rectangle region from Fig.~\ref{fig:slope1} compared to the total number of pairs in that bin. Again, at close angular separations, $5^{\prime\prime}<\theta<32^{\prime\prime}$ we see an excess of true pairs. The error bars show Poisson uncertainties.}
        \label{fig:ang_sep2}
\end{figure}

In the upper panel of Fig.~\ref{fig:slope1} we show the height and slope of the third bin from Fig.~\ref{fig:ang_sep1}, which is comprised by a large fraction of likely genuine binary systems. Here we observe that most points are distributed along a narrow stripe, therefore we use this region to define a rectangle with height > 0.4 and slope between 2.1 and 3.2. This narrow region contains pairs with \textit{isotachs} that lie parallel to the halo track in the RPM diagram. 
In principle, based on the \textit{isotach} criterion alone, these pairs could be regarded as good candidates.
However, this is not true for the last few bins of Fig.~\ref{fig:slope1}, 
where there is no information on true binaries, as the distribution of pairs in this region is consistent with random alignments.
Consequently, if we assume that these last two bins are pure contamination, we can use them to find the normalization of the random distribution, i.e. the zero-point of the dashed line in Fig.~\ref{fig:ang_sep2}.

\begin{figure}
    \centering
        \includegraphics[width=\columnwidth]{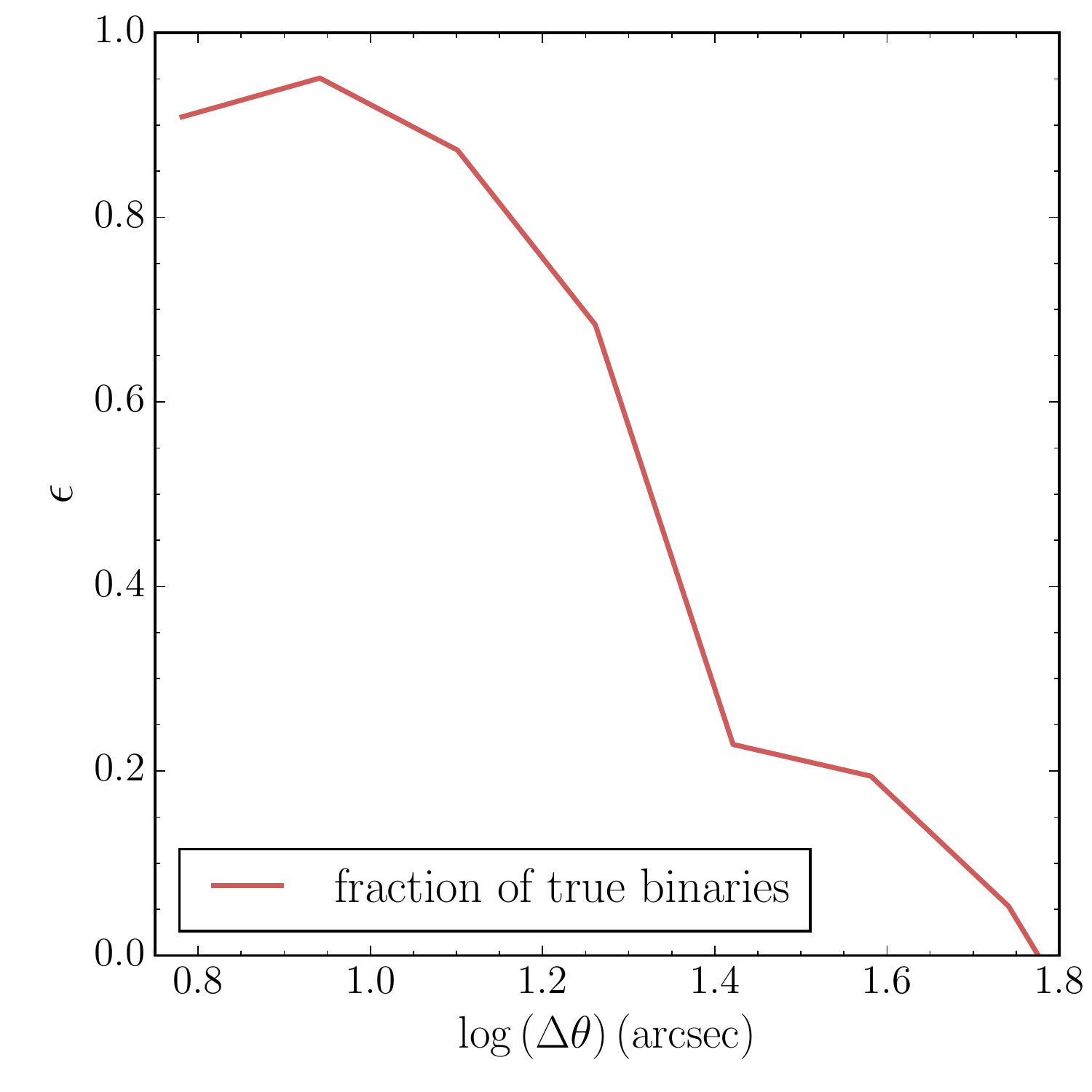}
        \caption{Fraction of true binaries, $\epsilon$ using bins from Fig.~\ref{fig:ang_sep2}. At separations larger than 63$^{\prime\prime}$ the fraction reaches 0, so no useful information can be obtained for larger separations.}
        \label{fig:frac_cont}
\end{figure}

In Fig.~\ref{fig:ang_sep2} we now see the distribution of angular separation for the pairs that fall into the enclosed region from the lower panel in Fig.~\ref{fig:slope1}. The dashed line shows again the number of random pairs, and they increase with angular separation. 
At separations less than 25$^{\prime\prime}$ there is a clear excess of true pairs. For separations between 25$^{\prime\prime}$ and 63$^{\prime\prime}$ there may be some excess of genuine binaries, but for separations larger than 63$^{\prime\prime}$ the pairs are consistent with random, chance alignments.
We then define the fraction of true binaries as $\epsilon$ = 1 - $n_{\rm obs}$/$n_{\rm rand}$ following \citet{ses08}, where $n_{\rm obs}$ corresponds to the number of observed pairs, and $n_{\rm rand}$ is the number given by the dashed line. 
We show this fraction as a function of the angular separation in Fig.~\ref{fig:frac_cont}.
Here we can see that $\epsilon$ decreases steeply from the fourth bin onwards, at separations larger than $\sim$ 20$^{\prime\prime}$. At separations larger than 63$^{\prime\prime}$ we see that this fraction reaches 0.

Finally, a subsample of these 5000 pairs was selected based on their observability from the southern hemisphere as the final catalog. This is composed of $\sim$400 halo wide binary candidates, and their positions in the RPM diagram can be seen in Fig.~\ref{fig:rpm_all_before}.

\section{Observations and data reduction}

For the candidates that passed our selection criteria we obtained long-slit spectra. Observations were carried with the COSMOS instrument at the Blanco telescope in Cerro Tololo Inter-American Observatory (CTIO) with the blue VPH grism covering a wavelength range of 5600-9600 \AA\,and a 0.9$^{\prime\prime}$ slit reaching a spectral resolution of $\sim$ 2100. Also with the EFOSC2 spectrograph at the NTT in La Silla with grism \#20 covering a wavelength range of 6047-7147 \AA\, and a 0.5$^{\prime\prime}$ slit reaching a spectral resolution of $\sim$ 3200. Finally with the KOSMOS instrument at the Kitt Peak National Observatory (KPNO) with the red VPH grism covering a wavelength range of 5600-9600 \AA\, and a 0.9$^{\prime\prime}$ slit reaching a spectral resolution of $\sim$ 2200. From the initial sample of 400 candidate pairs we obtained spectra for 88 pairs.

We highlight the fact that we do not need high resolution spectra given that we carefully select pairs that are close and move together in the sky. Therefore, we only need radial velocities within a few km/s in precision to confirm their association, given that the dispersion in the Galactic halo is of $\sim$ 140 km/s \citep{BH2016}.

For the lowest resolution spectrograph used (R$\sim$ 2100) the precision in radial velocity measurements is  $\Delta v_{r}\sim$ 15 km/s.

On the other hand, our sample is composed of halo stars which are faint targets with magnitudes in the range $14<V \lesssim 20$, hence long exposure times were necessary. We planned our observations aiming for a minimum signal-to-noise ratio of $S/N=10$, which is sufficient to resolve the general spectral features we need to estimate the radial velocities via cross-correlation techniques.

For the faintest targets ($V > 17$) combined exposures were obtained, taking 2 or 3 spectra per object depending on how faint the target was. In order to save time, for the cases where this was possible, stars of a pair were observed together placing both of them in the slit. This was done for the cases where angular separations were small enough in order to be able to observe them at the same time in the slit. Generally for cases where $\Delta\theta$ was larger than 70'' this was not possible. We always observed stars of a pair with the same instrument.

The process of data reduction involved all the standard procedures for cosmetics correction, extraction of the spectra and a final wavelength calibration which was done for each of the 88 pair candidates of wide binaries using \textit{ccdproc} and \textit{apextract} from \scriptsize{IRAF}. \normalsize. 

\section{Analysis}
In this section we will present the results from our radial velocity measurements for our catalog comprised of 88 candidate pairs. A full version of this catalog is available online. 

In Sec.~\ref{sec:binarity} we presented a method to analyse the contamination in our catalog due to chance alignments without including radial velocity information. Here we want to establish if the pairs that have small $\Delta v_{r}$ are consistent with being genuine binaries.  

\subsection{Radial velocity measurements}
\label{sec:radial_vel}
We measure the radial velocities of our sample using the \textit{xcsao,rvsao} \scriptsize{IRAF} \normalsize task, which is part of a package for obtaining radial velocities of stars and galaxies from optical spectra using the cross-correlation method developed by \citet{kurtz}.
The correlation procedure compares each spectrum with an input template. Given the importance of having templates with a high signal-to-noise ratio we use the PHOENIX library, which is a collection of synthetic spectra that covers effective temperatures from 2300 to 12000K and a wavelength range from 3000 to 25000 \AA\, with a high resolution of R $\approx$ 500,000 \citep{phoenix}.

We applied the conversion from vacuum to air wavelength to the synthetic spectra using the refractive index from \citet{edlen}.
%\newline

Given the low-resolution of our data, we lack detailed spectral information (i.e., precise spectral type and metallicity for each star), and thus we were not in a position to choose a single best template in advance for any given target. Therefore, we developed a procedure that involved the cross-correlation with the full library of spectra. However, as we selected only halo stars from the RPM diagram, we assume that our entire sample is composed of dwarfs. In consequence, we select templates with log g between 3.5 and 4.5 and [Fe/H] of -1.5 and effective temperatures in the range $2300~\lesssim~T_{\rm eff}~(\mbox{K})~\lesssim~8000$.

Once the cross-correlation with each template is done, we have to select a velocity from all the results obtained from this. Besides the velocity, \textit{xcsao} gives a reliability factor \textit{R} which depends on the amplitude of the cross-correlation peak, hence at larger values of $R$, the correlation is better. We plotted the obtained \textit{R} values from the cross-correlation as a function of the effective temperature of each template and adopted a Gaussian fit for this distribution, as can be seen in Fig.\ref{fig:R_value}. This allowed us to select a range of effective temperatures to consider (and for which the cross-correlation was better) and then, for the velocities that fell in that same range in effective temperature, the final velocity was taken from the median of all of these values. 

To determine how well the cross-correlation method is performing, we plot the error given by this method and its corresponding R value, which can be seen in Fig.\ref{fig:err_cross_corr}. For larger R's the error decreases until converging to the value of the spectral resolution of the spectrograph. Thus, choosing higher values of R also gives small errors in the cross-correlation.  

\begin{figure}
\centering
	\includegraphics[width=\columnwidth]{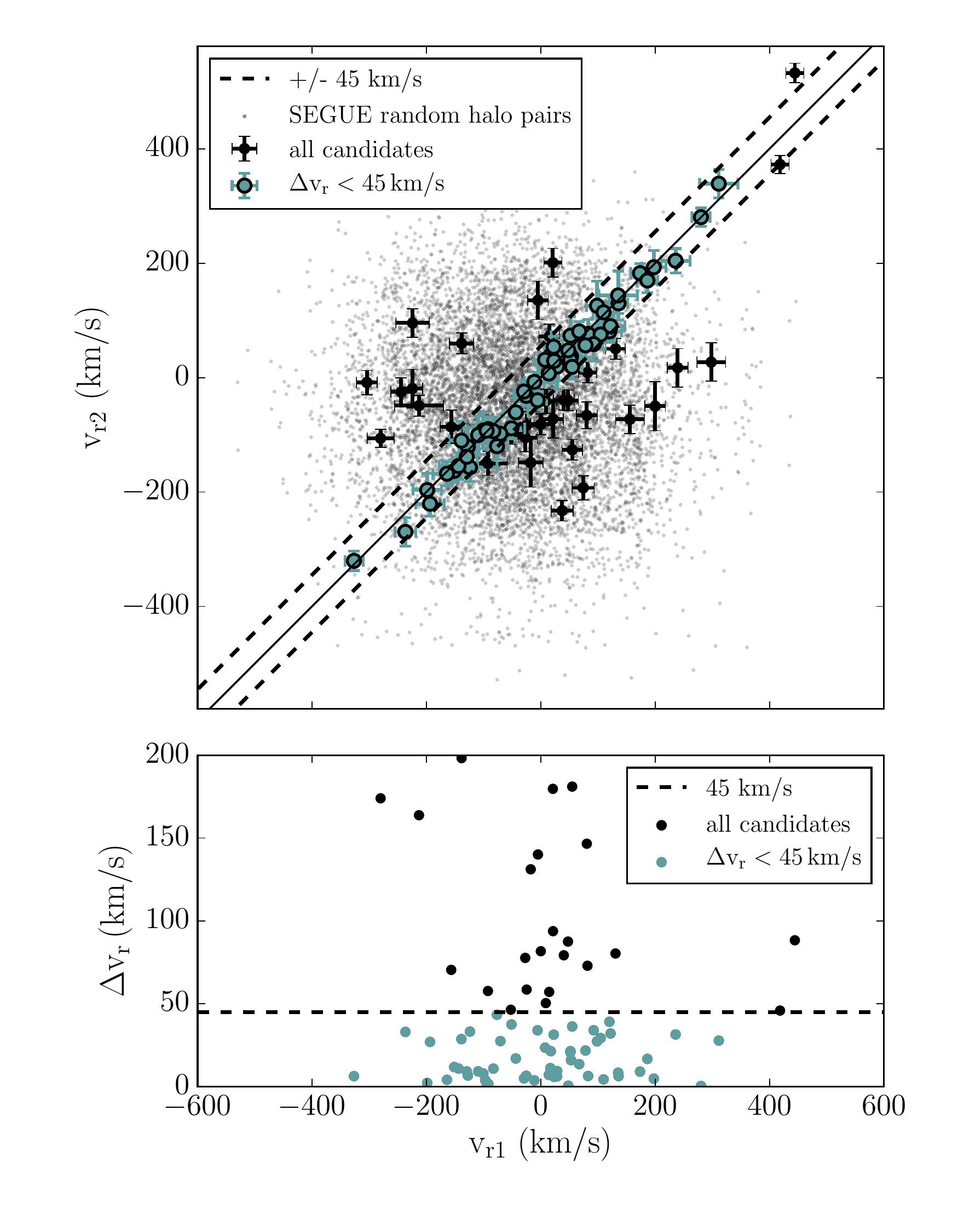}
    \caption{The upper and lower panels show the radial velocity measurements for all the candidates of our catalog. Each point in this plot corresponds to a pair. The central solid line corresponds to the 1-1 slope, and dashed lines show 3$\sigma$ of the precision in radial velocity from the instruments we used. The lower panel shows that the difference in radial velocity of our candidates is mostly concentrated in velocities less than 45 Km/s. $\Delta v_{r}>200$ km/s are omitted in the lower panel.}
    \label{fig:velocities}
\end{figure}

The result of all our radial velocity measurements for our 88 candidate pairs and their
corresponding uncertainties can be seen in Fig.~\ref{fig:velocities}. In this figure the velocity components of the primary versus the secondary star of the selected candidates can be seen in blue dots, therefore each point in this plot corresponds to a pair. The background is the result of a query in SEGUE for halo stars from the RPM diagram for which radial velocity measurements were available. Approximately 3000 stars with RV measurements were found, none of them belonged to our catalog. Then, in order to be able to plot them in Fig.~\ref{fig:velocities} a random selection of pairs for these stars was made. This was done considering that these velocities follow the velocity distribution of halo stars and assigning random velocities for a pair following this distribution. Note that the width of the random distribution of points, in both axes is consistent with the expected velocity dispersion of the Galactic halo which is is $\sim$ 140 km/s. 

It is expected that, because of low orbital velocities, wide binaries should have very similar proper motions and radial velocities, thus they should follow a 1-to-1 slope, which is the central solid line in Fig.~\ref{fig:velocities}. Most of the candidates follow this trend, also the candidates follow the distribution of velocities expected for halo stars, with $\sim$ 140 km/s of dispersion in velocity. Hence, we can say that the independent radial velocity vetting confirms that our criterion in proper motion alone already seems to select a large fraction of genuine halo wide binaries, given the velocity distribution of our candidates in this plot. However, we have to further analyse our sample in order to establish a quantitative result of \textit{binarity} for our candidate pairs in the catalog.

\begin{figure}
    \centering
        \includegraphics[width=\columnwidth]{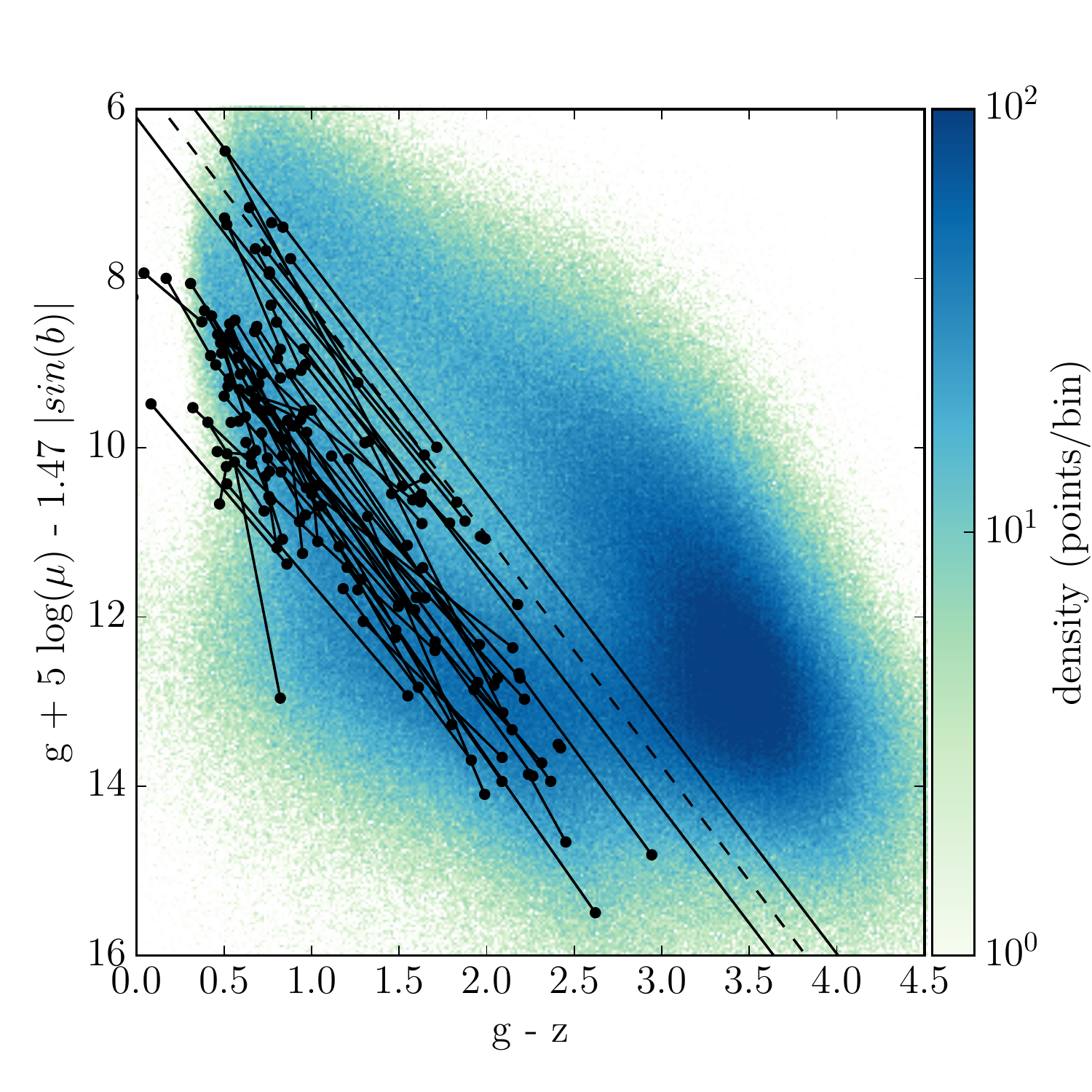}
        \caption{RPM diagram for our candidates with RV measurements in black dots connected by an \textit{isotach}, an isochrone vertically displaced in the RPM. We expect that real pairs are connected by an approximately parallel line with respect to the halo track since members of a binary system should have similar metallicities and essentially the same proper motion but with different luminosities.}
        \label{fig:rpm_conn}
\end{figure}

Besides our criteria in proper motion and angular separation, we did not apply any extra criterion on photometry besides the cut in $r<20$ to select the candidates for our catalog. Nonetheless, in Fig.~\ref{fig:rpm_conn} we see that if we connect our candidates with a line most of them lie parallel with respect to the halo track in the RPM diagram. This \textit{isotach} \citep{cg04} is an isochrone vertically displaced due to velocity (proper motion) in the RPM diagram. We should expect this behavior since both members of a binary should have similar metallicities and essentially the same proper motion but with different luminosities. Therefore, the line connecting their positions in the RPM diagram should be approximately parallel to the stellar halo track. There are a few cases in Fig.~\ref{fig:rpm_conn} where this is not the case, which can occur for two reasons: (1) either the wide binary is real but its photometry has been compromised, or (2) there is contamination of systems that are not real pairs, therefore we have to find a way to establish the amount of chance alignments in our catalog. Additionally, for the cases when the wide binary does not have a similar origin and it is dynamically captured, the metallicities are not necessarily similar. This could also explain why some of the lines are not parallel in this figure. However, \citet{andrews2} find that, for disk wide binaries, the metallicities of the components are identical to within less than 0.1 dex which most likely rules out dynamical capture as a major formation channel.

\subsection{Distances}
Our sample is composed of stars moving at relatively small proper motions ($\mu >$ 30 mas/yr) and of relatively faint stars covering the range in magnitudes $14<r<20$ therefore, we could think that these are distant stars. In order to actually estimate their distances we use the photometric parallax relation derived using globular clusters, where five globular clusters observed with SDSS with distances in the range 7-12 kpc were used to constrain the shape of  their color-magnitude relation $M_{r}^{0}(g-i)$, with $g-i<$ 0.8, by simultaneously fitting SDSS photometry data. We follow \citep{ivezic}, specifically their equations A2 and A3 to obtain the absolute magnitude for our candidates and equation A7 which is an extended photometric parallax relation, valid for 
$0.2 < g - i < 4.0$, given that our catalog of halo wide binary candidates covers a wide range in $g -i$. 

With the previous relation and setting the metallicity to -1.5 dex, which corresponds to the median value of the halo \citep{ivezic}, the absolute magnitude in the $r$ band ($M_{r}$) can be obtained as a function of color and metallicity. Unfortunately, not all of the candidates fall in the region of color $0.2 < g - i < 4.0$ and 8 pairs are left out, therefore no photometric distances can be calculated for these. We obtain a mean distance of $\sim$ 1 kpc for our candidates with RV measurements, with some of them extending up to 3.5 kpc.

\subsection{Candidates with RV measurements}

\begin{figure}
    \centering
        \includegraphics[width=\columnwidth]{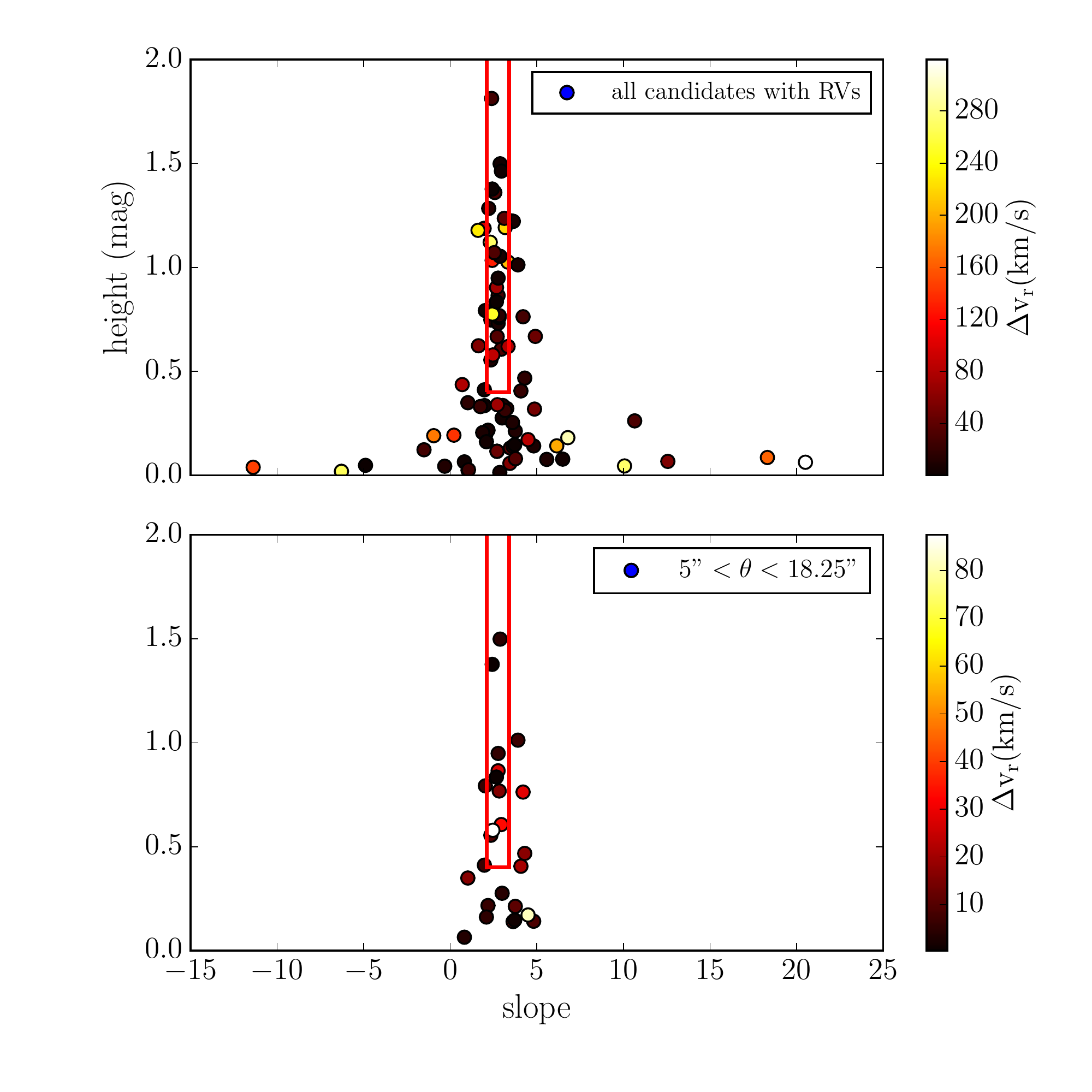}
        \caption{The upper panel shows the distribution of pairs for our entire catalog with RV measurements, with angular separations, $5^{\prime\prime}<\theta < 200^{\prime\prime}$. We see that in the region where we expect to have true binaries, shown by the red rectangle we have some contamination. In the lower panel we show our candidates with $\theta < 18.25^{\prime\prime}$ for which we find a 68\% probability of \textit{binarity}.}
        \label{fig:height_vels}
\end{figure}

Now that we have established a reliable method to quantify the degree of contamination, and to determine the fraction of true bounded systems, we proceed to test it by applying it to the candidates in our catalog with RV measurements. In Fig.~\ref{fig:height_vels} we show the height and slope from the RPM diagram for this sample, color coded by their difference in radial velocity. 
In the upper panel of this figure we see that there are some candidates with large $\Delta v_r$ inside the red rectangle. 
As we are including our entire catalog, which has pair with large angular separations, this result is consistent with our previous findings, where we expect to have significant contamination from random alignments. 
In the lower panel of this figure we show the subsample of pairs with $\theta<18.25^{\prime\prime}$, where we expect to have mostly genuine binaries, as can be seen from Fig.~\ref{fig:frac_cont}. 
We observe that less than $\sim$20\% of pairs have large differences in RV, implying much less contamination.
In consequence, these results are consistent with what we already obtained, and it shows the robustness of our previous analysis.  

\begin{figure}
    \centering
        \includegraphics[width=\columnwidth]{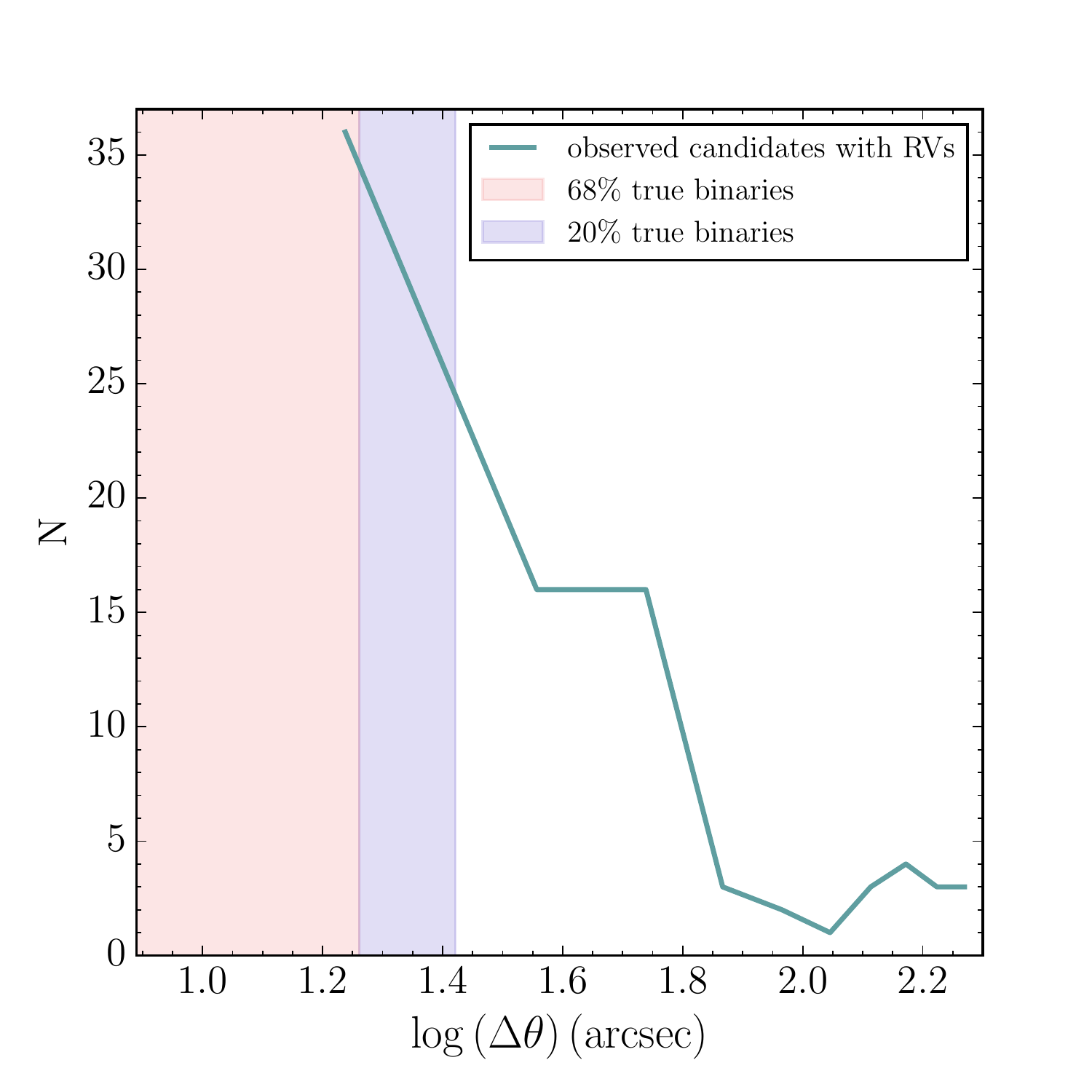}
        \caption{Fraction of true binaries, $\epsilon$ for our observed candidates with RV measurements. After separations larger than 18.25$^{\prime\prime}$ the fraction of true binaries decreases until it reaches 0 (see Fig.~\ref{fig:frac_cont}).}
        \label{fig:frac_cont2}
\end{figure}

We now proceed to determine the fraction of true binaries in our catalog with RVs using our proposed method. In Fig.~\ref{fig:frac_cont2} we show the distribution in angular separation for our candidates. The colored regions indicate the pairs that have up to 68\% and 20\% of confidence of being truly bounded in light pink and blue, respectively. These regions were determined based on the results shown in Fig.~\ref{fig:frac_cont}. From the distribution shown in Fig.~\ref{fig:frac_cont2}, we observe that at angular separations larger than 25$^{\prime\prime}$ we expect to have less than 20\% of true binaries.

\begin{figure}
    \centering
        \includegraphics[width=\columnwidth]{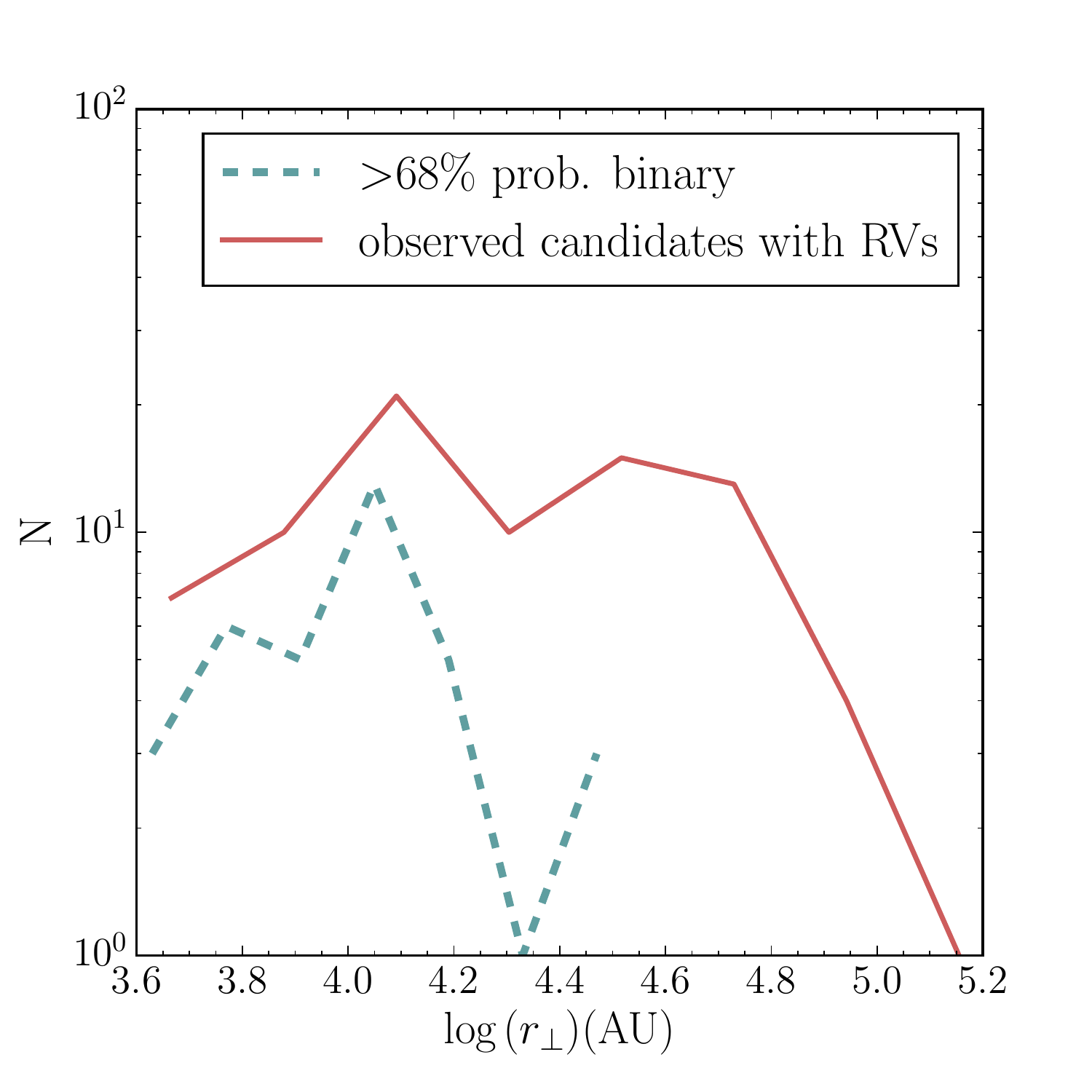}
        \caption{Distribution of projected distances in Astronomical Units (AU) for the entire catalog with RV measurements in light red and also in light blue the pairs in the bin for which we have confidence larger than 68\% of having genuine binaries. The pairs in light blue are the same ones from the lower panel of Fig.~\ref{fig:frac_cont2}.}
        \label{fig:prob_vels}
\end{figure}

With the calculated distances for our candidates we proceed to calculate their projected distances in the sky in Astronomical Units (AU). In Fig.~\ref{fig:prob_vels} we show the distribution of projected distance for our complete catalog of halo wide binary candidates with RV measurements, as well as the candidates that have 68\% or more probability of being genuine binaries. 
The entire sample has a roughly flat distribution up to $r_\perp\sim 6\times 10^4$ AU, where the number of pairs sharply drops.
Conversely, the candidates with large probabilities of being true binaries are concentrated at smaller distances, as most of the contamination comes from 
large projected distances. 

Our complete catalog with RV measurements reaches up to $\sim$ 0.9 parsecs in projected distance according to Fig.~\ref{fig:prob_vels}. However, once we analyse the contamination from chance alignments, this sample decreases and reaches only up to $\sim$ 0.1 parsecs in projected distance. We obtain for this subsample a 68\% probability or more of \textit{binarity}. For larger separations we have no reliable information on genuine pairs, as shown by Fig.~\ref{fig:ang_sep2}. 

\begin{figure}
    \centering
        \includegraphics[width=\columnwidth]{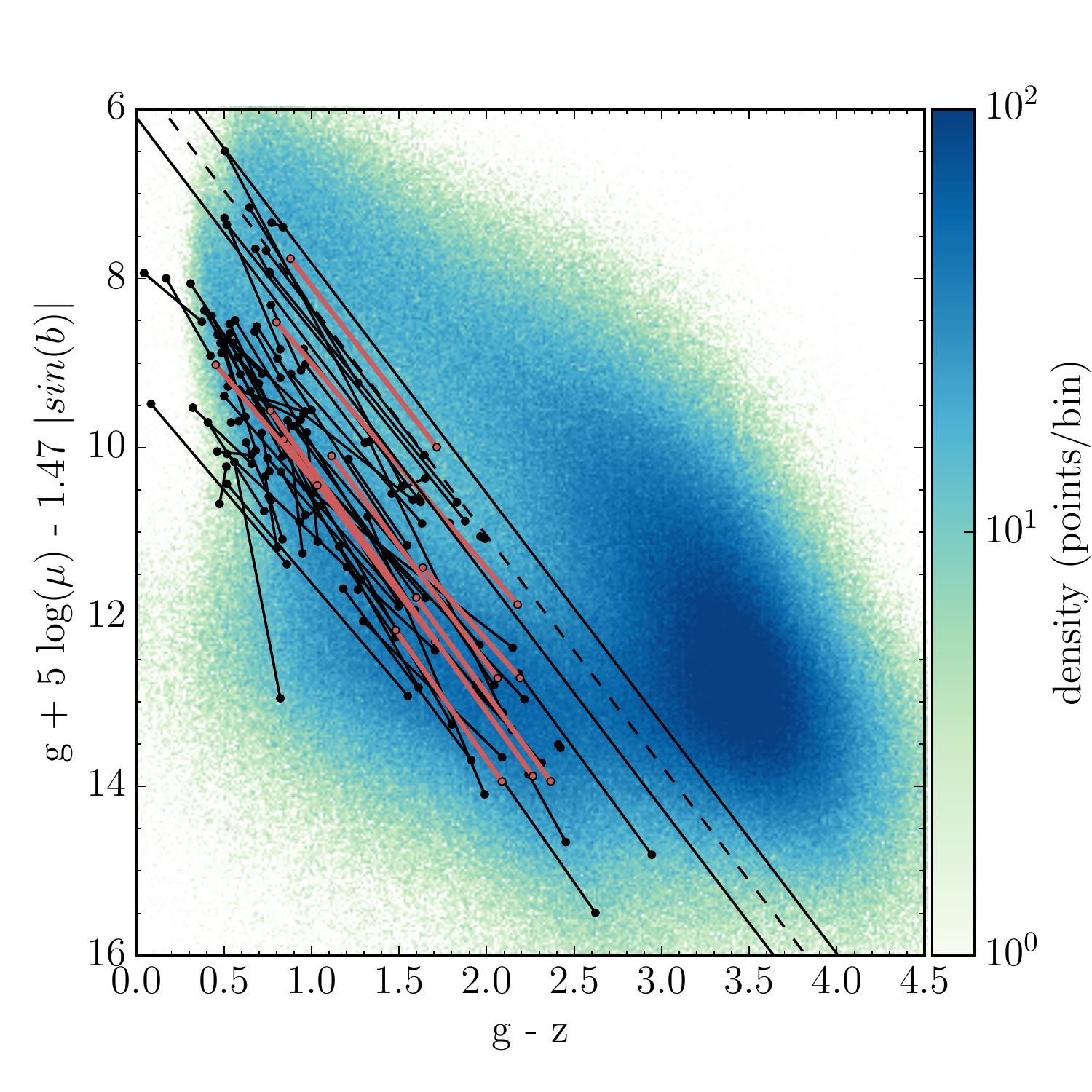}
        \caption{RPM diagram showing the distribution of pairs for our entire catalog with RV measurements, with angular separations, $5^{\prime\prime}<\theta < 200^{\prime\prime}$ in black dots. In pink we show the candidates with $\theta < 18.25^{\prime\prime}$ for which we find a 68\% of probability of \textit{binarity}.}
        \label{fig:rpm_4}
\end{figure}

Finally, for the pairs with $\theta < 18.25^{\prime\prime}$ for which we find a 68\% of probability of \textit{binarity} we show their connecting \textit{isotachs} in the RPM diagram. We present these pairs in Fig.~\ref{fig:rpm_4} connected by pink lines. Clearly our criterion selects the well connected pairs in this diagram. For comparison, we show the entire sample with RVs connected by black lines.

\subsection{Selection effects as a function of separation}

\begin{figure}
\includegraphics[width=\columnwidth]{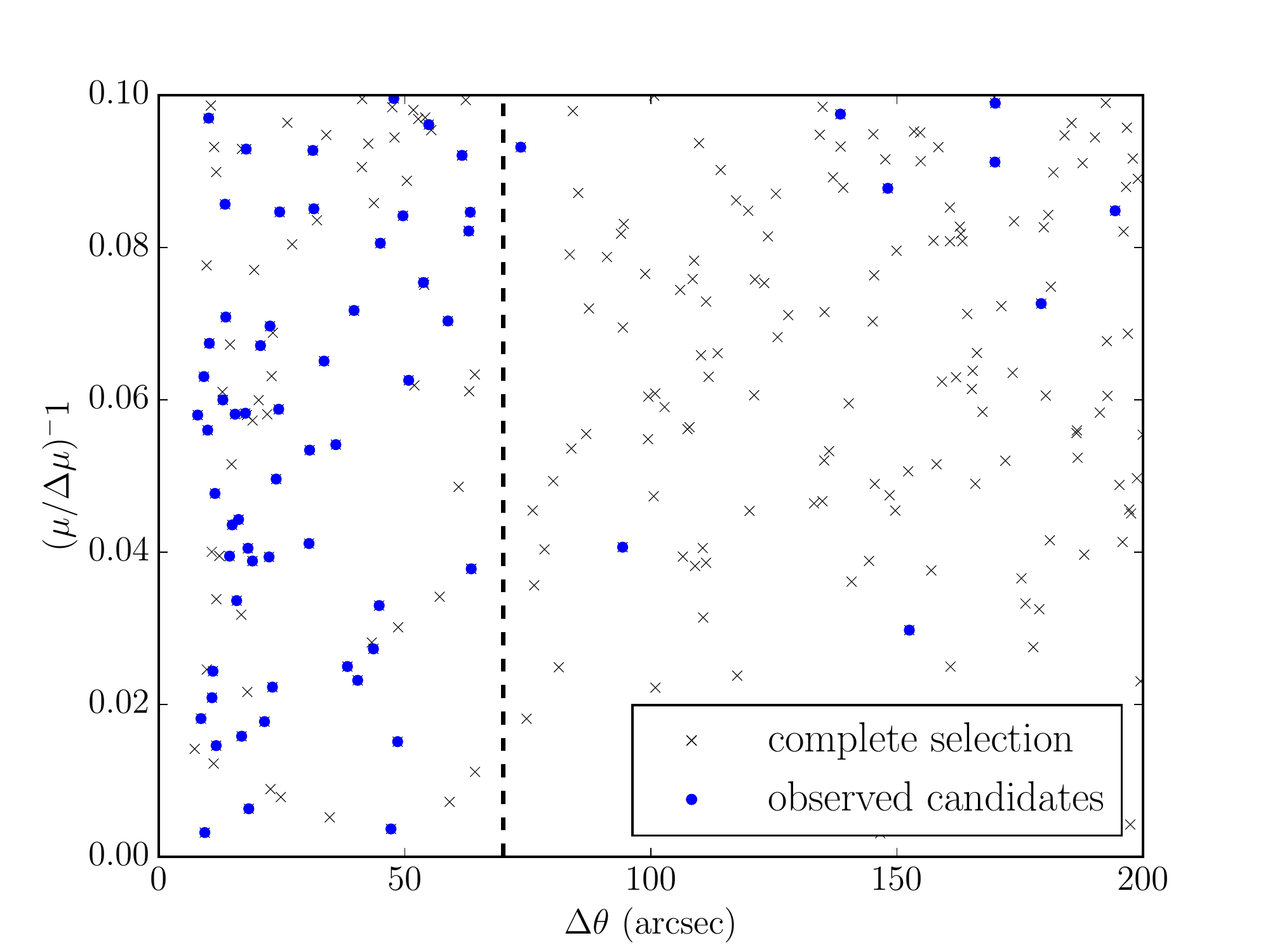}
 \caption{1/($\mu$/$\Delta\mu$) criterion vs the angular separation for the candidates between 5$^{\prime\prime}$$<\Delta\theta< $200$^{\prime\prime}$, $\delta_{1}$  and $\delta_{2} <  25$ deg, $V_{1}$ and $V_{2} \lesssim$ 19. The dashed line shows the limit of 70'' in angular separation, where our sample is complete.}
\label{fig:mu_deltamu}
\end{figure}

If this catalog is intended to be used to improve constraints on dark matter in the halo of the Galaxy in the form of MACHOs, we need to have a complete selection of candidates covering a certain parameter range to draw conclusions that will be statistically significant from the sample. Our selection criterion $\mu$/$\Delta\mu$ is substantial and already eliminates contamination from pairs that are chance alignments. Moreover, it is a selection criterion independent of angular separation, crucial for dynamical population studies. In Fig.~\ref{fig:mu_deltamu} we plot the quantity 1/$(\mu/\Delta\mu)$, where smaller values imply that the criterion is stronger and that the halo wide binary candidate has more chance of being an actual bound system. After applying our selection criteria, the sample we assembled is still fairly large consisting of $\sim$ 400 pairs, as we mentioned in Sec.~\ref{sec:sel_pairs}. However, if we consider further constraints, we can have a more or less complete observed sample of halo wide binaries. In Fig.~\ref{fig:mu_deltamu} we show the candidates that could be observed from the southern hemisphere, and with magnitudes for the primary and secondary star, $V_{1}$ and $V_{2} \lesssim$ 19. From this selection, our final radial velocity-vetted sample is shown as blue opened circles, and black crosses are the candidate halo wide binaries that could not be observed due to weather conditions. We note in this figure that up to angular separations of 70$^{\prime\prime}$ we observed most of the pairs in that region.

\section{Conclusions}
We report a new method to find genuine wide binaries from the halo of our Galaxy. We also report a new assembled catalog of halo wide binaries using photometry and proper motion information from SDSS, complemented by ground-based radial velocities from medium-resolution spectra. Wide binaries have low binding energies and large collisional cross sections, which makes them easily disrupted by encounters with different perturbers (MACHOs being one of them). Therefore, one of their applications is that they can be used to place constrains on the amount of halo dark matter composed of these compact objects.
With this new catalog we expand the current small and local sample of genuine halo wide binaries (until today, best represented by that of CG04) and therefore we produce a new one probing the Galactic halo to distances never reached before with these objects, and free of selection effects as a function of separation. 

Because there are a large number of false pairs present in such a large sample of stars with similar proper motions, radial velocity measurements were obtained in order to confirm which of the candidates selected with proper motion criteria were genuine binaries. From the radial velocity measurements it was possible to test the adopted criterion in proper motion to select the halo wide binary candidates, and according to our results $\sim$ 68\% of the sample with separations up to $\theta<18.25^{\prime\prime}$ is composed of true bound systems, having consistent proper motions and consistent velocities. From the results obtained with the radial velocity measurements, we show that proper motion alone already selects many genuine halo wide binaries at close angular separations.

Moving groups and clusters could be misclassified as wide binaries given that the velocity difference of the group members should be low. However, such associations span over many square degrees on the sky and are found with projected separations > 1 pc \citep{andrews1}. Because our sample is composed of stars with projected separations < 1 pc, we do not expect to have any significant contamination from members of moving groups.

We find pairs up to $\sim$ 0.9 parsecs in projected distances, but once we analyse the contamination fraction, we find that the sample reaches only up to 0.1 parsecs. For larger separations we cannot reliably establish a fraction of true bounded systems. 
Future work includes the calculation of Galactic orbits for the candidate pairs, given that all the information necessary for them is available in this catalog (proper motions, radial velocities, positions and photometric distances). Finally, this new assembled catalog will certainly help in the future goal of statistically selecting even larger samples from the entire SDSS for example, as well as from \textit{Gaia} and Panstarrs.

\section*{Acknowledgements}
The authors would like to thank the anonymous referee for very useful comments that improved a first version of this paper. We thank Felipe G. Goicovic for extensive and invaluable discussions. We also thank F. van de Voort for providing material that was useful at the beginning of this project. J.C and J.Ch acknowledge support from Proyecto FONDECYT Regular 1130373. J.C. also  acknowledges support from the SFB 881 program (A3) and the International Max Planck Research School for Astronomy and Cosmic Physics at Heidelberg University (IMPRS-HD). J.Ch acknowledges support from BASAL PFB-06 Centro de Astronom\'ia y Tecnolog\'ias Afines; and by the Chilean Ministry for the Economy, Development, and Tourism's Programa Iniciativa Cient\'ifica Milenio grant IC 120009, awarded to the Millennium Institute of Astrophysics. This work has made use of data from the European
Space Agency (ESA) mission Gaia (http://www.cosmos.esa.int/gaia),
processed by the Gaia Data Processing and Analysis Consortium (DPAC,
http://www.cosmos.esa.int/web/gaia/dpac/consortium). Funding for the DPAC
has been provided by national institutions, in particular the institutions participating
in the Gaia Multilateral Agreement.

%%%%%%%%%%%%%%%%%%%%%%%%%%%%%%%%%%%%%%%%%%%%%%%%%%



\bibliographystyle{mnras}
\bibliography{example} % if your bibtex file is called example.bib

%%%%%%%%%%%%%%%%%%%%%%%%%%%%%%%%%%%%%%%%%%%%%%%%%%

%%%%%%%%%%%%%%%%% APPENDICES %%%%%%%%%%%%%%%%%%%%%

\appendix
\section{Radial velocities: template fitting}
\begin{figure}
\centering
	\includegraphics[width=\columnwidth]{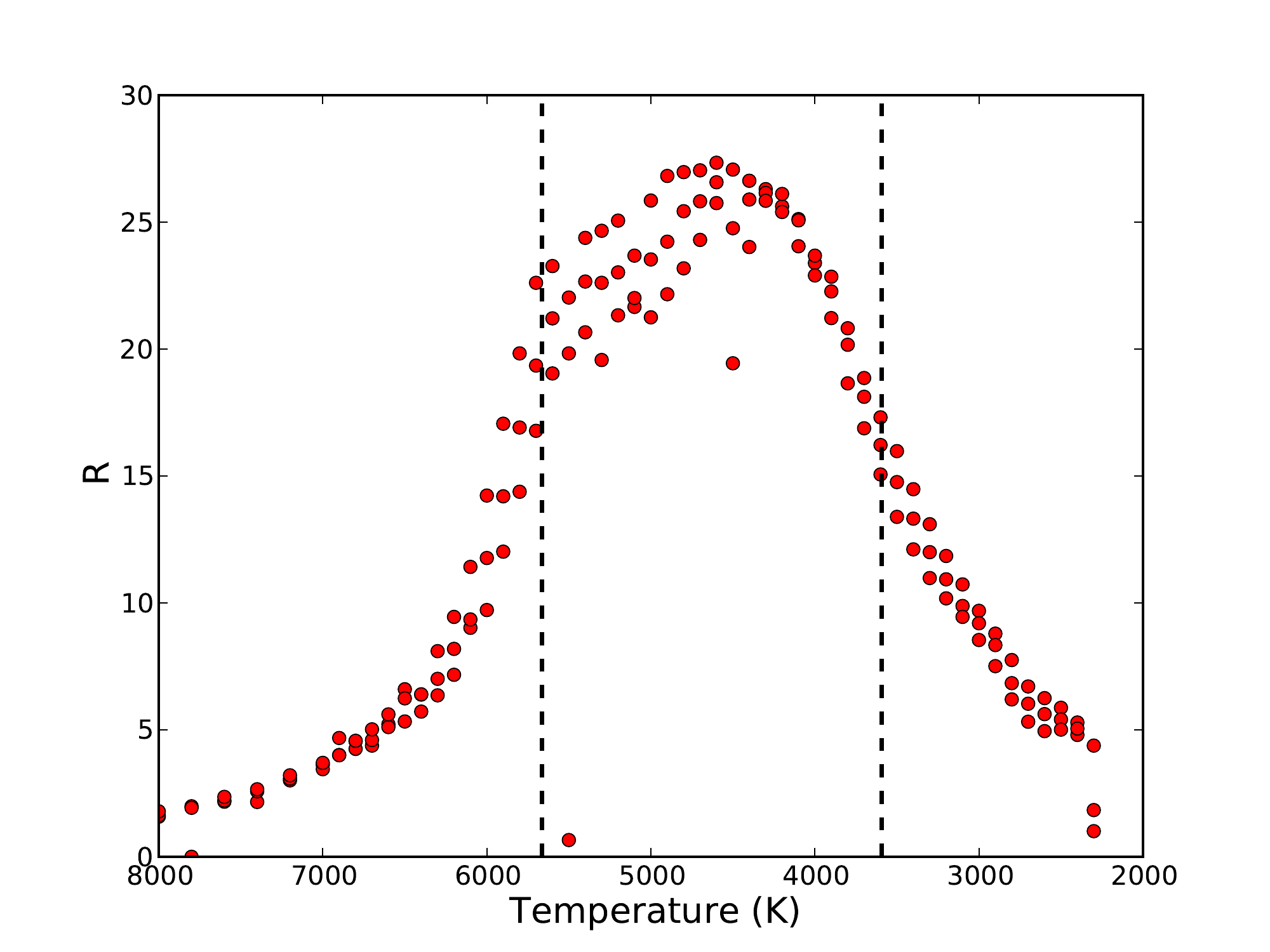}
    \caption{Example of our template-fitting procedure for radial velocity measurements. Here we show for one arbitrary target, the correlation R values as a function of the temperature of each template. Each point is the result of the cross-correlation between the spectra of one candidate star of a pair with the template. Dashed lines correspond to 1$\sigma$ of the Gaussian profile fitted to the points.}
    \label{fig:R_value}
\end{figure}

\begin{figure}
\centering
	\includegraphics[width=\columnwidth]{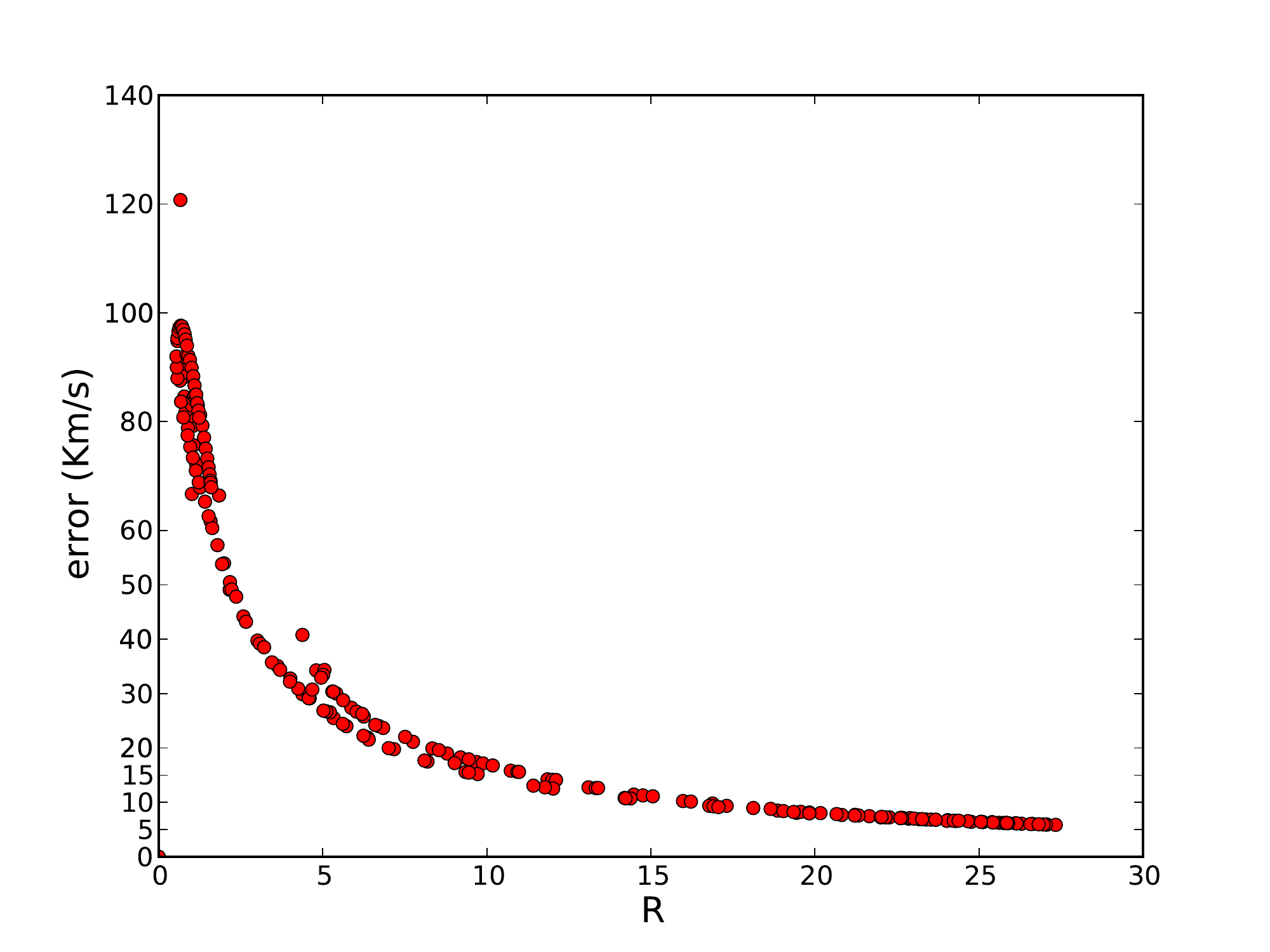}
    \caption{Error of the cross-correlation (km/s) as a function of R. Each point corresponds to the result of the cross-correlation between the spectra of one candidate star of a pair with a template.}
    \label{fig:err_cross_corr}
\end{figure}

In this section we present some of our results of the cross-correlation method we used to calculate the radial velocities in our catalog. In Fig.~\ref{fig:R_value} and Fig.~\ref{fig:err_cross_corr} each red dot in the plot is the result of the correlation of the spectrum of the star and a different template. As was noted in 
Sec.\ref{sec:radial_vel}, a cross-correlation with the complete range of templates covering 3.5 $<$log g $<$4.5 is necessary because we do not have the stellar parameters of the wide halo binary candidates. The synthetic spectra have effective temperature information, which is used to plot Fig.~\ref{fig:R_value}. This is the first step in the
determination of the velocity. Besides the velocity, \textit{xcsao} gives a reliability factor \textit{R} which depends on the amplitude of the cross-correlation peak, hence at larger values of $R$, the correlation is better. We expect the candidates to be dwarfs because of their location in the RPM diagram their effective temperatures should be around $2300 \lesssim T_{eff}\,(\mbox{K})\, \lesssim 8000$ just as Fig.~\ref{fig:R_value} shows, and we adopt a Gaussian fit to select the best $R$ values. The errors obtained by the cross-correlation are illustrated in Fig.~\ref{fig:err_cross_corr}, where we observe that for larger values of $R$ the error decreases until converging into $\sim$ 5 km/s. Finally, once it has been established the range of $R$ and temperature, we consider which velocities fall into this range, and this is how we select the velocity for each candidate pair.

\section{\textit{Gaia} DR2 validation}

\begin{figure}
\centering
	\includegraphics[width=\columnwidth]{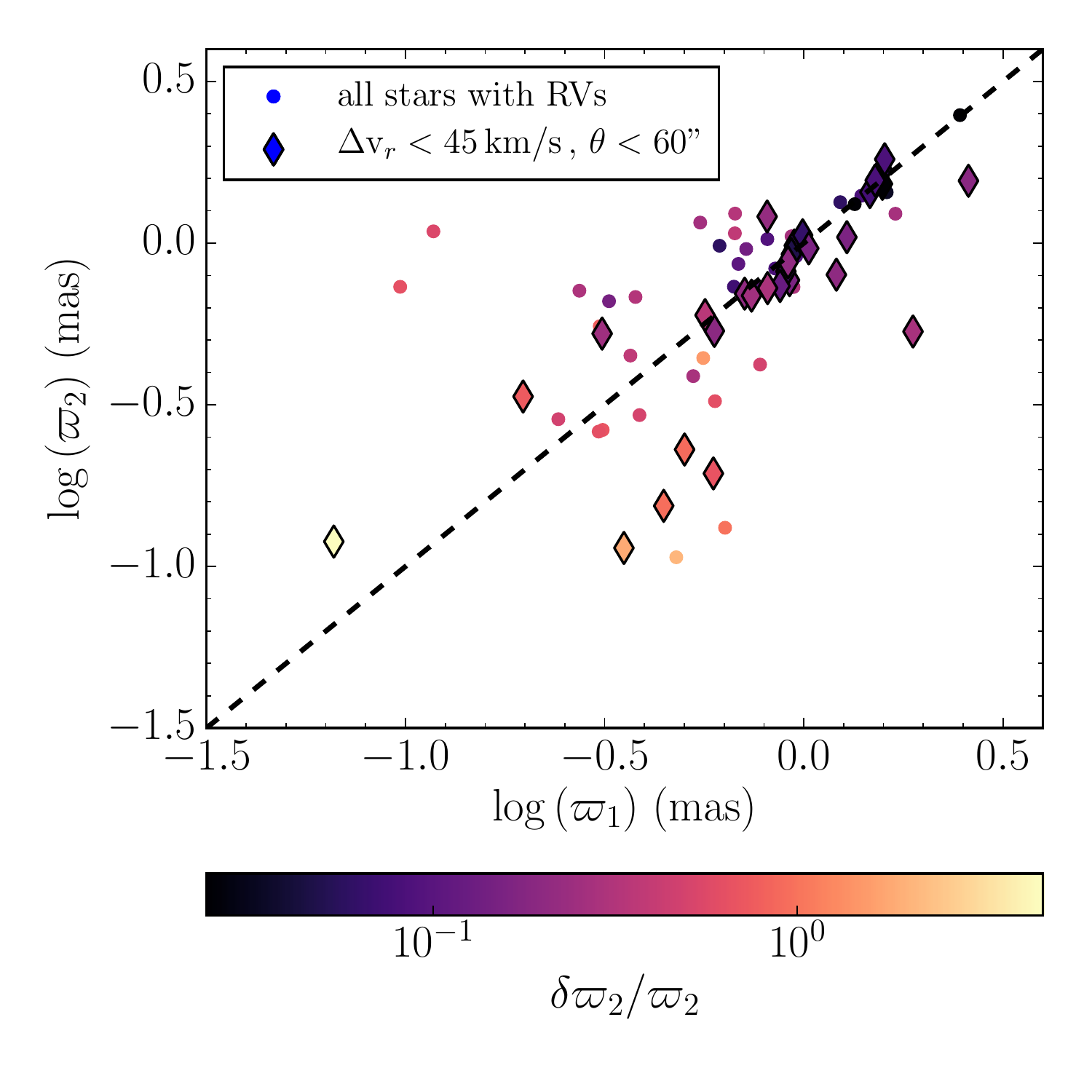}
    \caption{The dots show \textit{Gaia} DR2 parallaxes for the 74 candidate pairs with radial velocity measurements color coded by the parallax uncertainty of the secondary star ($\delta\varpi_{2}$/$\varpi_{2}$), which is dominant. The diamonds are the pairs that have $\Delta v_{r} < 45$ km/s and separations $\theta < 60''$. }
   \label{fig:gaia_dr2_1}
\end{figure}

\begin{figure}
\centering
	\includegraphics[width=\columnwidth]{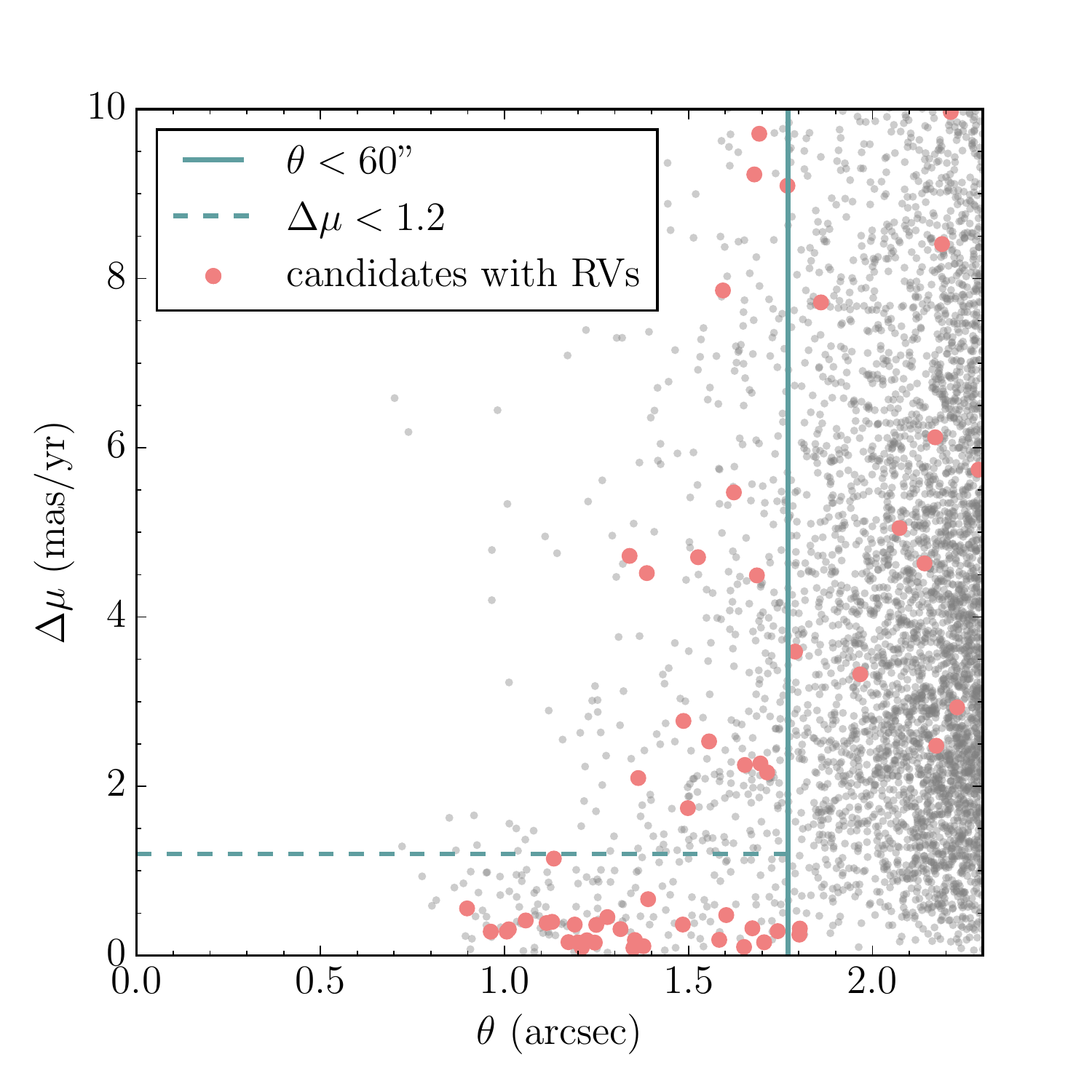}
    \caption{\textit{Gaia} DR2 proper motion difference for the 74 candidate pairs with RVs and their angular separation. We observe that most of them fall into the region with small differences in proper motions $\Delta\mu<1.2$.} 
    \label{fig:gaia_dr2_2}
\end{figure}

\begin{figure}
\centering
	\includegraphics[width=\columnwidth]{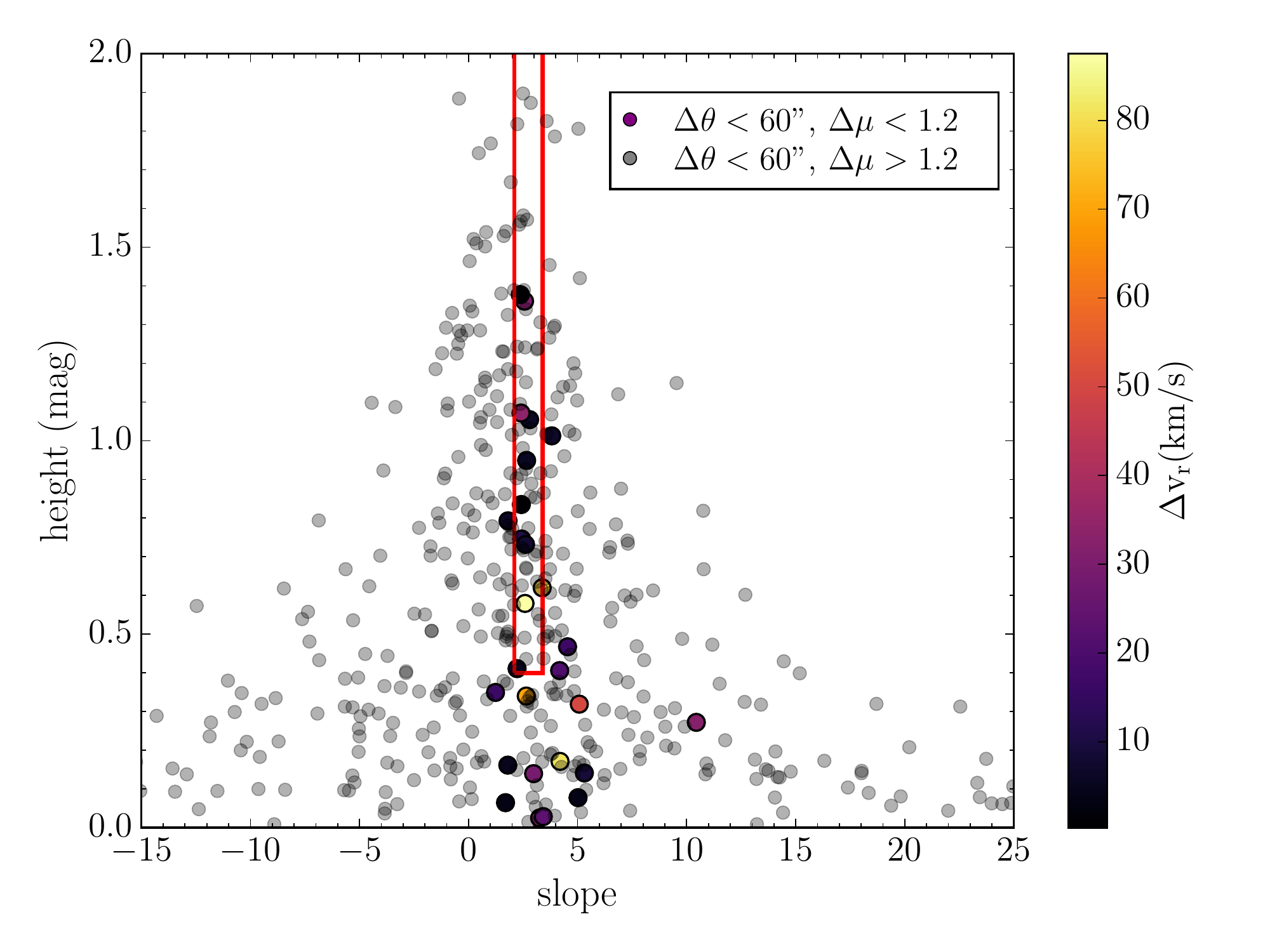}
    \caption{Height and slope of the \textit{isotach} connecting candidate pairs in the RPM diagram, for pairs satisfying $\theta <60''$ and $\Delta\mu < 1.2$ color coded by their RV difference. In black dots we show all of the candidates from the initial catalog of $\sim$ 5,000 pairs with no RVs, that satisfy $\theta <60''$ and $\Delta\mu > 1.2$. The red rectangle shows the region of this plot where we should have well behaved pairs connected by a parallel line with respect to the halo track in the RPM.}
    \label{fig:gaia_dr2_3}
\end{figure}

This work was completed and submitted prior to \textit{Gaia's} DR2. However, it is of interest to see how the results and methods presented in this paper look like in the light of \textit{Gaia}. Therefore, we present Figures~\ref{fig:gaia_dr2_1},~\ref{fig:gaia_dr2_2} and~\ref{fig:gaia_dr2_3} to illustrate this. We do not undertake any further and extensive analysis because this is beyond the scope of this paper and we use \textit{Gaia} data only as validation of our results. 

From the 88 candidate pairs with RVs presented in this paper we found 74 pairs in \textit{Gaia} DR2 with improved proper motions and parallaxes. In Fig.~\ref{fig:gaia_dr2_1} we show the parallaxes of the primary and the secondary star color coded by the parallax uncertainty of the secondary star ($\delta\varpi_{2}$/$\varpi_{2}$) which is the dominant error. A large fraction of our candidates at large parallaxes (closer distances) have small parallax uncertainties ($\delta\varpi_{2} < 10\%$) and these show common parallaxes in the large majority of cases. Additionally, for $\varpi >$ 0.6 mas, all but two of our candidates with $\Delta$v$_{r}$ $<$ 45 km/s and $\theta <60"$ (diamonds) are consistent with the 1-1 line, which constitutes independent and strong validation of our selection method. Below $\varpi$ $\sim$ 0.6 mas, candidates with consistent radial velocities deviate from the 1-1 line, but these are all stars with poor \textit{Gaia} DR2 parallaxes, so the test cannot be trusted in this regime.

Furthermore, we illustrate in Fig.~\ref{fig:gaia_dr2_2} the difference in proper motions updated by \textit{Gaia} DR2 for our candidate pairs and their angular separation. We see that most of the sample is located at small $\Delta\mu$, where a large fraction of them are likely binaries. We then proceed to select a region within $\Delta\mu < 1.2$ and at separations $\theta < 60"$.

In Fig.~\ref{fig:gaia_dr2_3} we show their distribution in the height and slope diagram, analogous to what we presented in Fig.~\ref{fig:height_vels}. We show the candidate pairs with $\theta <60''$ and $\Delta\mu < 1.2$, that fall in the enclosed region in Fig.~\ref{fig:gaia_dr2_2}. For comparison, we also show pairs that satisfy $\theta <60''$ but with large differences in proper motions, $\Delta\mu > 1.2$ from the initial catalog of $\sim$ 5,000 pairs (with no measured RVs).

Fig.~\ref{fig:gaia_dr2_3} shows that the boxed region in the slope-height diagram is a good way to select candidates, but even at $\theta < 60''$
it leaves substantial contamination that must be vetted in some other way, either with additional proper motion epochs or radial velocity measurements.

\section{Catalog of Halo Wide Binaries}

In this section we present the catalog description of our halo wide binary candidates. The complete table is available online.

\begin{table}
\caption{Catalog description. Each row in the table corresponds to a pair.} % title of Table
\centering % used for centering table
\begin{tabular}{ccc} % centered columns (4 columns)
\hline\hline %inserts double horizontal lines
Column & Units & Description\\ 
\hline % inserts single horizontal line
\verb ID &  &  Identifier for each star (integer) \\
\verb ra1 &  deg & Right ascension; star 1 \\
\verb dec1 &  deg & Declination; star 1 \\
\verb ra2 &  deg & Right ascension; star 2\\
\verb dec2 &  deg & Declination; star 2 \\
\verb pml1   & mas yr$^{-1}$ & Proper motion in galactic longitude;  \\
&& star 1\\
\verb pml2 & mas yr$^{-1}$ & Proper motion in galactic longitude;\\ &&star 2  \\
\verb pmb1 & mas yr$^{-1}$  & Proper motion in galactic latitude; \\ &&star 1  \\
\verb pmb2 & mas yr$^{-1}$ &  Proper motion in galactic latitude;\\ && star 2  \\
\verb psfMag_r1 & mag & Psf magnitude in the r band; star 1 \\
\verb psfMag_r2 & mag & Psf magnitude in the r band; star 2 \\
\verb ang_sep & arcseconds & Angular separation of the pair\\
\verb Vel1 & km s$^{-1}$ & Radial velocity; star 1 \\
\verb err1 & km s$^{-1}$ & Error in the radial velocity; star 1 \\
\verb Vel2 & km s$^{-1}$ & Radial velocity; star 2 \\
\verb err2 & km s$^{-1}$ & Error in the radial velocity; star 2 \\

\hline %inserts single line
\end{tabular}
\hfill\hfill\hfill\hfill
\label{table:nonlin} % is used to refer this table in the text
\end{table}
%\end{landscape}

%

%%%%%%%%%%%%%%%%%%%%%%%%%%%%%%%%%%%%%%%%%%%%%%%%%%

% Don't change these lines
\bsp	% typesetting comment
\label{lastpage}
\end{document}